\documentclass[preprint,amsmath,amssymb,aps,nofootinbib]{revtex4-2}

\usepackage{graphicx}
\usepackage{caption}
\usepackage{subcaption}
\usepackage{dcolumn}
\usepackage{bm}
\usepackage{physics}
\usepackage{xcolor}
\usepackage{filecontents}
\usepackage{soul}
\usepackage{hyperref}

\allowdisplaybreaks

\newcommand{\dbar}{d\hspace*{-0.08em}\bar{}\hspace*{0.1em}}
\newcommand{\deltabar}{\delta\hspace*{-0.2em}\bar{}\hspace*{0.1em}}
\newcommand{\lb}{\left(}
\newcommand{\rb}{\right)}
\newcommand{\lsb}{\left[}
\newcommand{\rsb}{\right]}

\begin{document}

\title{Stochastic parameters for scalar fields in de Sitter spacetime}
\author{Archie Cable}
 \email{archie.cable18@imperial.ac.uk}
\author{Arttu Rajantie}
 \email{a.rajantie@imperial.ac.uk}
\affiliation{Department of Physics,\\Imperial College London,\\London, SW7 2AZ, United Kingdom}

\date{\today}

\begin{abstract}
The stochastic effective theory approach, often called stochastic inflation, is widely used in cosmology to describe scalar field dynamics during inflation. The existing formulations are, however, more qualitative than quantitative because the connection to the underlying quantum field theory (QFT) has not been properly established. A concrete sign of this is that the QFT parameters depend on the renormalisation scale,  and therefore the relation between the QFT and stochastic theory must have explicit scale dependence that cancels it. In this paper we achieve that by determining the parameters of the second-order stochastic effective theory of light scalar fields in de Sitter to 
linear order in the self-coupling constant $\lambda$. This is done by computing equal-time two-point correlators to one-loop order both in QFT using dimensional regularisation and the $\overline{\rm MS}$ renormalisation scheme and the equal-time four-point correlator to leading order in both theories, and demanding that the results obtained in the two theories agree. With these parameters, the effective theory is valid when $m\lesssim H$ and $\lambda^2\ll m^4/H^4$, and therefore it is applicable in cases where neither perturbation theory nor any previously proposed stochastic effective theories are.
\end{abstract}

\maketitle

\tableofcontents

\section{Introduction}
\label{sec:introduction}

The mathematical framework that one uses to describe the dynamics of inflation is scalar quantum field theory (QFT) in de Sitter spacetime \cite{Birrell-Davies_book,tagirov:1973,chernikov:1968,bunch-davies:1978,Starobinsky:1980,guth_inflation,linde_inflation,slow_roll_liddle,Vazquez_Gonzalez_2020_slowroll}. Specifically, the long-distance behaviour of scalar fields is directly related to inflationary observables \cite{baumann_book}. Spectator fields that existed during inflation can become observationally relevant in the present day and thus warrant further study. Examples of their application include curvature and isocurvature perturbations \cite{Sasaki:1986, Linde:1997, Leach:2001}, dark matter generation \cite{Peebles_1999_darkmatter, Hu_2000_darkmatter, Markkanen_2018_darkmatter,Jukko:2021} and primordial black hole abundance \cite{Lyth2005,Cable:2023_PBH}, electroweak vacuum decay \cite{Espinosa_2008_vdecay,Herranen_2014_vdecay,Markkanen_2018_vdecay,Camargo-Molina:2022p1,Camargo-Molina:2022p2} and gravitational-wave background anisotropy \cite{gwb_anisotropy}. 

Scalar QFT cannot be solved exactly so one has to turn to approximate methods. For example, when one considers a quartic self-interacting theory, with coupling $\lambda$, of a scalar field with non-zero mass $m$, the standard approach is to perform a perturbative expansion in small $\lambda$ about the free field solution. However, such a perturbative expansion is only valid in the regime $\lambda\ll m^4/H^4$, where $H$ is the Hubble parameter, because the expansion fails to converge beyond this limit \cite{allen_folacci_perturbative_corr, allen_perturbative_corr, Sasaki:1993, Suzuki:1994}. This is an infrared (IR) problem and so is particularly prevalent when one is interested in the long-distance dynamics of the fields, and so one must look for alternative methods \cite{hu_oconnor_symm_behaviour,boyanovsky_quantum_correct_SR,Serreau_2011,Arai_2012,Gautier_2013,Herranen_2014,Gautier_2015,Guilleux_2015,Boyanovsky:2015,Guilleux:2016,Boyanovsky:2016,Nacir:2019}. 

The focus of this paper will be one such method: the stochastic effective theory of the long-distance behaviour of scalar fields in de Sitter. The premise is that, for sufficiently light fields $m\lesssim H$, long wavelength field modes are stretched by the expanding spacetime to such a degree that they can be considered classical. The remaining short wavelength modes remain quantum, but their contribution to the long-wavelength dynamics can be summarised by a statistical noise contribution. This method was pioneered by Starobinsky and Yokoyama \cite{starobinsky:1986,Starobinsky-Yokoyama:1994}, who derived stochastic equations from the slow-roll ``overdamped'' (OD) equations of motion that govern the inflaton using a cut-off method to separate the long and short wavelength modes. Stochastic inflation has since become a powerful tool for performing computations relating to the inflaton \cite{ Tsamis_2005,Finelli_2009,Finelli_2010,Vennin_2015,Grain_2017,Hardwick_2017_post-inflation,Tokuda_2018,Tokuda_2018_2,Glavan_2018,Cruces_2019,Firouzjahi_2019,Pinol_2019,Hardwick_2019,Pattison_2019,Moreau_2020,Moreau:2020gib,Rigopoulos_2016,Moss_2017,Prokopec_2018,bounakis2020feynman,garbrecht_rigopoulos:2014,Garbrecht_2015_Fdiag,morikawa_1990,rigopoulos2013fluctuationdissipation,Levasseur_2013,Rigopoulos_2016,Moss_2017,Pinol_2019, Pinol:2020,Moreau_2020,Moreau:2020gib,Wilkins:2021,Andersen_2022}. However, due to the nature of the slow-roll equations, it is limited to the regime $m\ll H$ and $\lambda\ll m^2/H^2$. These conditions are satisfied if the scalar field in question is the inflaton existing in slow-roll, but there are plenty of scenarios where one might wish to go beyond this regime, particularly if one is interested in studying spectator fields \cite{Markkanen_2018_darkmatter,Markkanen:2019,Markkanen_2020,Jukko:2021,Cable:2023_PBH}.

This can be achieved with a second-order stochastic effective theory~\cite{Grain_2017,Cruces_2019,Pattison_2019,Cable:2021,Cable:2022}, in which case the stochastic dynamics takes place in phase space, and no slow-roll assumption is needed. However, in the existing literature on both first-order and second-order stochastic theories, the parameters of this stochastic effective theory have not been computed beyond tree-level. In particular, the mass parameter $m_S^2$ of the stochastic theory has been taken to be equal to the renormalised mass parameter $m_R^2$ of the quantum field theory. Obviously, this cannot be accurate because $m_R^2$ depends on the arbitrary renormalisation scale $M$, and there is nothing in the stochastic theory that can cancel this dependence. This introduces a relative error of order $O(\lambda H^2/m^2)$ in the stochastic theory, limiting its range of validity and meaning that it cannot be used to compute precise quantative predictions.

In this paper we extend the approach used in Refs.~\cite{Cable:2021,Cable:2022}, to compute the parameters of the stochastic effective theory to full one-loop order in perturbation theory. This calculation does not suffer from the infrared problem to the same extent as direct perturbative calculations of observables, and it gives a relation between the two theories that cancels the renormalisation scale dependence explicitly. Instead of using a cut-off approach to derive our stochastic equations, we consider second-order stochastic equations that resemble the classical equations of motion for a scalar field in de Sitter, but with general stochastic parameters: the stochastic mass $m_S$, stochastic coupling $\lambda_S$ and the noise contributions $\sigma_{ij}^2$. One can apply standard techniques to solve these equations to compute stochastic correlation functions, which can be compared with perturbative QFT correlators to determine what the stochastic parameters should be for the stochastic theory to be promoted to an effective theory of scalar fields in de Sitter. This method was first introduced for free fields in Ref. \cite{Cable:2021} before being extended to include self-interactions in Ref. \cite{Cable:2022}. 

In Ref. \cite{Cable:2022}, the matching between the stochastic theory and perturbative QFT was done by comparing the equivalent 2-point functions. On the perturbative QFT side, this required us to renormalise the theory because the $\mathcal{O}(\lambda)$ correction to the Feynman propagator is ultraviolet (UV) divergent. In Ref. \cite{Cable:2022}, we only considered the correction term to leading order such that $\lambda\ll m^4/H^4$, which meant we could neglect a detailed discussion of renormalisation schemes. We found that the stochastic parameters required to reproduce such a term were simply equal to those found in Ref. \cite{Cable:2021} for free fields. The result was a stochastic effective theory that was valid in the regime $m\lesssim H$ and $\lambda\ll m^2/H^2$. However, this doesn't give a full account of the $\mathcal{O}(\lambda)$ contributions to the stochastic parameters. In this paper, we will perform a more detailed analysis of the UV renormalisation on the QFT side, which can then be matched by stochastic results to give full expressions for the stochastic parameters to $\mathcal{O}(\lambda)$. Additionally, we will also consider the connected 4-point function in both approximations such that we can compute the relation between the stochastic coupling $\lambda_S$ and its QFT counterpart $\lambda$. Thus, we will extend the regime of our stochastic theory to $m\lesssim H$ and $\lambda^2\ll m^4/H^4$.

We start by giving a full account of perturbative QFT in Sec. \ref{sec:scalar_QFT}. We introduce the free field 2-point functions in Sec. \ref{subsec:free_QFT} before outlining the in-in path integral formalism in Sec. \ref{subsec:in-in_formalism}. We then perform the UV renormalisation of the 2-point function using the $\overline{\text{MS}}$ scheme of dimensional regularisation in Sec. \ref{subsec:2-pt_function_one-loop}, with further details of the calculation in Appendix \ref{app:dim_reg_de_Sitter}. This amounts to a mass redefinition in order to absorb the UV divergent terms. As a result of dimensional regularisation, the mass parameter becomes dependent on the renormalisation scale $M$. To round out the QFT, we compute the connected 4-point function in Sec. \ref{subsec:4-pt_functions}. In Sec. \ref{sec:second-order_theory}, we move onto the second-order stochastic theory. This follows the same procedure as in Ref. \cite{Cable:2022}, except now the matching procedure is performed using the fully UV-renormalised QFT results. We also include the computation of the stochastic connected 4-point function in Sec. \ref{subsubsec:4-pt_2o_stochastic_correlator}, which is compared with its QFT counterpart to find an expression for $\lambda_S$. Finally, we perform a comparison between results from perturbative QFT, OD stochastic theory and second-order stochastic theory, including both the new results from this paper and the old results from Ref. \cite{Cable:2022}. Additionally, we discuss how the $M$-dependence of the stochastic parameters becomes important. We conclude with some remarks in Sec. \ref{sec:conclusion}.

\section{Quantum field theory in de Sitter spacetime}
\label{sec:scalar_QFT}

We begin by considering a perturbative approach to scalar QFT in de Sitter spacetime. We will consider a spectator scalar field $\phi(t,\mathbf{x})$ with a scalar potential $V(\phi)$ in a de Sitter background parametersed by the scale factor 
\begin{equation}
\label{scale_factor}
a=e^{Ht}=-\frac{1}{H\eta},
\end{equation}
where $H=\frac{1}{a}\dv{a}{t}$ is the Hubble parameter, which is constant, and $t$ and $\eta$ are physical and conformal time respectively. They are related by
\begin{equation}
    \label{conformal-->physical}
    \eta=-\frac{1}{H}e^{-Ht}.
\end{equation}
The action for scalar fields in de Sitter is given by
\begin{equation}
    \label{scalar_action}
    S[\phi]=\int d^4x a(t)^3\lsb\frac{1}{2}\Dot{\phi^2}-\frac{1}{2}\frac{\lb\nabla_{\mathbf{x}}\phi\rb^2}{a(t)^2}-V(\phi)\rsb.
\end{equation}
Introducing the field momentum $\pi(t,\mathbf{x})$, the equations of motion are given by 
\begin{equation}
    \label{scalar_eom}
    \begin{pmatrix}\Dot{\phi}\\\Dot{\pi}\end{pmatrix}
    =
    \begin{pmatrix}\pi\\-3H\pi-\frac{\nabla_\mathbf{x}^2\phi}{a(t)^2}-V'(\phi)\end{pmatrix},
\end{equation}
where primes and dots denote derivatives with respect to $\phi$ and $t$ respectively. Throughout this paper, we will consider a field with mass $m$ and a quartic self-interaction parameterised by $\lambda$ such that the potential is $V(\phi)=\frac{1}{2}m^2\phi^2+\frac{1}{4}\lambda\phi^4$. Additionally, we include a non-minimial coupling to gravity $\xi$, which is absorbed into the mass term such that $m^2=m_0^2+12\xi H^2$.

\subsection{Free scalar QFT}
\label{subsec:free_QFT}

We will first consider free fields such that $\lambda=0$. Following standard procedures of second quantisation \cite{Birrell-Davies_book,bunch-davies:1978,tagirov:1973,chernikov:1968}, the mode functions in the Bunch-Davies vacuum are given by
\begin{subequations}
\label{mode_functions}
    \begin{align}
    \label{phi_k_mode}
    \phi_k(t)&=\sqrt{\frac{\pi}{4Ha(t)^3}}\mathcal{H}^{(1)}_\nu\lb\frac{k}{a(t)H}\rb,\\
    \label{pi_k_mode}
    \pi_k(t)&=-\sqrt{\frac{\pi}{16Ha(t)^3}}\lsb3H\mathcal{H}^{(1)}_\nu\lb\frac{k}{a(t)H}\rb+\frac{k}{a(t)}\lb\mathcal{H}^{(1)}_{\nu-1}\lb\frac{k}{a(t)H}\rb-\mathcal{H}_{\nu+1}^{(1)}\lb\frac{k}{a(t)H}\rb\rb\rsb,
    \end{align}
\end{subequations}
where $k=\abs{\mathbf{k}}$, $\nu=\sqrt{9/4-m^2/H^2}$ and $\mathcal{H}_\nu^{(1)}(z)$ are the Hankel functions of the first kind. Note that there is an equivalent set of solutions using the Hankel functions of the second kind $\mathcal{H}_\nu^{(2)}(z)$.

The quantities we are interested in are correlation functions: expectation values of the Bunch-Davies vacuum state. The most basic of these is the scalar field two-point function, which is computed in $\mathbf{k}$-space as
\begin{equation}
    \label{k-space_2-pt_func}
    \begin{split}
        \bra{0}\hat{\phi}(t,\mathbf{k})\hat{\phi}(t',\mathbf{k}')\ket{0}=&\phi_k(t)\phi_{k'}^*(t')
        \\=&
        \frac{\pi}{4Ha(t)^{3/2}a(t')^{3/2}}\mathcal{H}^{(1)}_\nu\lb\frac{k}{a(t)H}\rb\mathcal{H}^{(2)}_\nu\lb\frac{k'}{a(t')H}\rb.
    \end{split}
\end{equation}
We can perform the Fourier transform to obtain the 2-pt function in coordinate space as
\begin{equation}
    \label{2-pt_func_k-space}
    \bra{0}\hat{\phi}(t,\mathbf{x})\hat{\phi}(t',\mathbf{x}')\ket{0}=\int\dbar^3\mathbf{k}e^{-i\mathbf{k}\cdot(\mathbf{x}-\mathbf{x}')}\bra{0}\hat{\phi}(t,\mathbf{k})\hat{\phi}(t',\mathbf{k})\ket{0}.
\end{equation}
Computing this integral \cite{chernikov:1968,tagirov:1973,Dowker:1976,bunch-davies:1978,Schomblond:1976} results in the positive (+) and negative (-) frequency Wightman functions in the Bunch-Davies vacuum
\begin{subequations}
    \label{wightman_functions_pre_contour}
    \begin{align}
        \begin{split}
        \Delta^+(x,x')&:=\bra{0}\hat{\phi}(t,\mathbf{x})\hat{\phi}(t',\mathbf{x}')\ket{0}
        \\&=\frac{H^2}{16\pi^2}\Gamma(\alpha)\Gamma(\beta)_2F_1\lb\beta,\alpha,2;1+\frac{\lb\eta-\eta'-i\epsilon\rb^2-\abs{\mathbf{x}-\mathbf{x}'}^2}{4\eta\eta'}\rb,
        \end{split}\\
        \begin{split}
        \Delta^-(x,x')&:=\bra{0}\hat{\phi}(t',\mathbf{x}')\hat{\phi}(t,\mathbf{x})\ket{0}
        \\&=\frac{H^2}{16\pi^2}\Gamma(\alpha)\Gamma(\beta)_2F_1\lb\beta,\alpha,2;1+\frac{\lb\eta-\eta'+i\epsilon\rb^2-\abs{\mathbf{x}-\mathbf{x}'}^2}{4\eta\eta'}\rb,    
        \end{split}
    \end{align}
\end{subequations}
where $\Gamma(z)$ is the Euler-Gamma function, $_2F_1(a,b,c;z)$ is the hypergeometric function and $\alpha=3/2-\nu$, $\beta=3/2+\nu$. The $i\epsilon$ prescription indicates the pole about which we perform our contour integration in the complex plane. Note that I have used conformal time $\eta$ here, defined in Eq. (\ref{conformal-->physical}), for convenience. From the equation of motion (\ref{scalar_eom}), the Wightman functions obey the equation
\begin{equation}
    \label{Wightman_funcs_eom}
    \lb\Box_{dS}+m^2\mp i\epsilon\rb \Delta^{\pm}(x,x')=0,
\end{equation}
where $\Box_{dS}=\partial_t^2+3H\partial_t-\frac{\nabla_\mathbf{x}^2}{a(t)^2}$. 

Physical correlators must be invariant under the de Sitter group. This means that the behaviour of such correlators can be written purely in terms of a de Sitter invariant combination of the spacetime coordinates. The quantity in question is
\begin{equation}
    \label{de_Sitter_invariant}
    \begin{split}
        y(x,x')&=\frac{\lb\eta-\eta'\rb^2-\abs{\mathbf{x}-\mathbf{x}'}^2}{2\eta\eta'}\\
        &=\cosh\lb H(t-t')\rb-\frac{H^2}{2}e^{H(t+t')}\abs{\mathbf{x}-\mathbf{x}'}^2-1.
    \end{split}
\end{equation}
We can write the Wightman functions in terms of the de Sitter invariant by expanding about small $\epsilon$ to give
\begin{subequations}
    \label{wightman_functions}
    \begin{align}
        \Delta^+(x,x')=&\frac{H^2}{16\pi^2}\Gamma(\alpha)\Gamma(\beta)_2F_1\lb\beta,\alpha,2;1+\frac{y}{2}\rb+\frac{iH^2}{32\pi}(4\nu^2-1)_2F_1\lb\beta,\alpha,2;-\frac{y}{2}\rb\theta(y)\theta(t'-t),\\    
        \Delta^-(x,x')=&\frac{H^2}{16\pi^2}\Gamma(\alpha)\Gamma(\beta)_2F_1\lb\beta,\alpha,2;1+\frac{y}{2}\rb+\frac{iH^2}{32\pi}(4\nu^2-1)_2F_1\lb\beta,\alpha,2;-\frac{y}{2}\rb\theta(y)\theta(t-t'),
    \end{align}
\end{subequations}
where we have introduced the Heaviside function $\theta(z)$. Henceforth, I will drop the $i\epsilon$ prescription. From the Wightman functions, we can build our other scalar 2-point correlators. For convenience, we define
\begin{subequations}
    \label{A&B_2-pt_amplitudes}
    \begin{align}
        A(y)&=\frac{H^2}{16\pi^2}\Gamma(\alpha)\Gamma(\beta)_2F_1\lb\beta,\alpha,2;1+\frac{y}{2}\rb,\\
        B(y)&=\frac{H^2}{32\pi}(4\nu^2-1)_2F_1\lb\beta,\alpha,2;-\frac{y}{2}\rb\theta(y).
    \end{align}
\end{subequations}
Then, we define various other 2-point functions in Table \ref{tab:2-pt_functions}.

\begin{table}[ht]
\begin{small}
    \centering
    \begin{tabular}{c|c|c|c|c}
        Name & Symbol & Correlator & Form & $\Box_{dS}+m^2=$\\ \hline &&&&\\
        Hadamard & $\Delta^H(x,x')$ & $\bra{0}\acomm{\hat{\phi}(t,\mathbf{x})}{\hat{\phi}(t',\mathbf{x}')}\ket{0}$ & $2A(y)+iB(y)$ & 0\\&&&&\\
        Causal & $\Delta^C(x,x')$ & $-i\bra{0}\comm{\hat{\phi(t,\mathbf{x})}}{\hat{\phi}(t',\mathbf{x}')}\ket{0}$ & $B(y)\lb\theta(t'-t)-\theta(t-t')\rb$ & 0 \\&&&&\\
        Advanced & $\Delta^A(x,x')$ & $-i\theta(t'-t)\bra{0}\comm{\hat{\phi}(t,\mathbf{x})}{\hat{\phi}(t',\mathbf{x}')}\ket{0}$ & $B(y)\theta(t'-t)$ & $\frac{1}{a(t)^3}\delta^{(4)}(x-x')$ \\&&&&\\
        Retarded & $\Delta^R(x,x')$ & $i\theta(t-t')\bra{0}\comm{\hat{\phi}(t,\mathbf{x})}{\hat{\phi}(t',\mathbf{x}')}\ket{0}$ & $B(y)\theta(t-t')$ & $\frac{1}{a(t)^3}\delta^{(4)}(x-x')$ \\&&&&\\
        Feynman & $i\Delta^F(x,x')$ & $\bra{0}T\hat{\phi}(t,\mathbf{x})\hat{\phi}(t',\mathbf{x}')\ket{0}$ & $A(y)$ & -$\frac{i}{a(t)^3}\delta^{(4)}(x-x')$ \\&&&&\\
        Dyson & $i\Delta^D(x,x')$ & $\bra{0}\Tilde{T}\hat{\phi}(t,\mathbf{x})\hat{\phi}(t',\mathbf{x}')\ket{0}$ & $A(y)+iB(y)$ & $\frac{i}{a(t)^3}\delta^{(4)}(x-x')$ 
    \end{tabular}
    \caption{The 2-point functions for free fields, built out of the Wightman functions (\ref{wightman_functions}).}
    \label{tab:2-pt_functions}
\end{small}
\end{table}

The most relevant 2-point function for this paper is the time-ordered function, called the Feynman propagator for free fields. It is given by
\begin{equation}
    \label{free_feynman_propagator}
    i\Delta^F(x,x'):=\bra{0}T\hat{\phi}(t,\mathbf{x})\hat{\phi}(t',\mathbf{x}')\ket{0}=\frac{H^2}{16\pi^2}\Gamma(\alpha)\Gamma(\beta)_2F_1\lb\beta,\alpha,2;1+\frac{y}{2}\rb,
\end{equation}
where $T(\Tilde{T})$ is the (anti-)time ordered operator. It obeys the equation
\begin{equation}
    \label{Feynman_prop_eom}
    \lb\Box_{dS}+m^2\rb i\Delta^F(x,x')=-\frac{i}{a(t)^3}\delta^{(4)}(x-x').
\end{equation}
One can study the leading IR behaviour of the Feynman propagator by considering the asymptotic expansion about large $y$. The leading terms in such an expansion are\footnote{Note that for $m>\sqrt{2}H$, there are other terms in the asymptotic expansion that dominate more than the second term in Eq. (\ref{free_feynman_long-distance_leading}). For this paper, we are interested in light fields $m\lesssim H$ so this will never be the case.}
\begin{equation}
    \label{free_feynman_long-distance_leading}
    i\Delta^F(x,x')=\frac{H^2}{16\pi^2}\lsb\frac{\Gamma(2\nu)\Gamma(\frac{3}{2}-\nu)}{\Gamma(\frac{1}{2}+\nu)}\qty(-\frac{y}{2})^{-\frac{3}{2}+\nu}+\frac{\Gamma(-2\nu)\Gamma(\frac{3}{2}+\nu)}{\Gamma(\frac{1}{2}-\nu)}\qty(-\frac{y}{2})^{-\frac{3}{2}-\nu}\rsb+...
\end{equation}
The leading IR behaviour of the timelike (equal-space) Feynman propagator is then given by
\begin{equation}
    \label{equal-space_phiphi_corr}
    \begin{split}
        \bra{0}T\hat{\phi}(t,\mathbf{x})\hat{\phi}(t',\mathbf{x})\ket{0}=&\frac{H^2}{16\pi^2}\frac{\Gamma(2\nu)\Gamma\qty(\alpha)(-4)^\alpha}{\Gamma\qty(\frac{1}{2}+\nu)}e^{-\alpha H(t-t')}\\&+\frac{H^2}{16\pi^2}\frac{\Gamma(-2\nu)\Gamma\qty(\beta)(-4)^\beta}{\Gamma\qty(\frac{1}{2}-\nu)}e^{-\beta H(t-t')}+...
    \end{split}
\end{equation}

\subsection{The in-in formalism}
\label{subsec:in-in_formalism}

Having laid the foundations with free fields, we now turn our attention to the more interesting situation where we include interactions such that $\lambda\ne0$. The addition of the interaction means we cannot straightforwardly compute the scalar correlators; the equations of motion (\ref{scalar_eom}) cannot be solved analytically. Instead, we will consider a perturbative approach where one performs a small-$\lambda$ expansion about the free solution that we computed in the last section. 

We will use the \textit{in-in}\footnote{Also known as Schwinger-Keldysh or closed time path} path integral formalism, where we begin in some initial vacuum state - the in state $\ket{0_-}$ - evolve our system to some intermediate state before evolving back to the in state \cite{Calzetta:1987,Calzetta:1988,Calzetta:1989,Calzetta_txtbk:2008}. The in-in generating functional is defined by
\begin{equation}
    \label{in-in_generating_functional_def}
    \begin{split}
    Z[J^+,J^-]=&_{J^-}\braket{0_-}{0_-}_{J^+}
    \\=&\int D\psi\bra{0_-}\Tilde{T}e^{i\int^{-\infty}_{t_*}dt\int d^3\mathbf{x}\sqrt{-g}J^-(x)\hat{\phi}(x)}\ket{\psi}\\&\times\bra{\psi}T e^{i\int^{t_*}_{-\infty}dt\int d^3\mathbf{x}\sqrt{-g}J^+(x)\hat{\phi}(x)}\ket{0_-},
    \end{split}
\end{equation}
where $J^{\pm}$ are external sources that source the evolution from/to $\ket{0_-}$ respectively, and we take the normalisation $Z[J,J]=1$. In the path integral representation, we introduce two auxiliary fields, $\phi^+$ and $\phi^-$, to differentiate the contributions from the two paths. The in-in generating functional is then given by
\begin{equation}
    \label{in-in_generating_functional_path_integral}
    Z[J^+,J^-]=\int D\phi^+\int D\phi^-e^{i\lb S^+[\phi^+]+J_x^+\phi_x^+-S^-[\phi^-]-J_x^-\phi_x^-\rb},
\end{equation}
where $S^\pm[\phi^\pm]=S[\phi^\pm]$ with the $\pm i\epsilon$ prescription, with $S[\phi]$ defined in Eq. (\ref{scalar_action}). Further, we have introduce the de Witt condensed notation for convenience, where repeated indices represent integrals over the spacetime coordinate: for example, $J_x\phi_x=\int d^4x \sqrt{-g(x)}J(x)\phi(x)$. All scalar correlators can be built from the in-in generating functional via
\begin{equation}
    \label{in-in_scalar_correlators}
    \begin{split}
    \bra{0_-}\Tilde{T}\lsb\hat{\phi}(x_1)...\hat{\phi}(x_n)\rsb& T\lsb\hat{\phi}(x'_1)...\hat{\phi}(x'_m)\rsb\ket{0_-}\\&=(-i)^{n-m}\frac{\delta^{n+m}Z[J^+,J^-]}{\delta J^-(x_1)...\delta J^-(x_n)\delta J^+(x'_1)...\delta J^+(x'_m)}\eval_{J^\pm=0}.
    \end{split}
\end{equation}
The Bunch-Davies vacuum states considered in Sec. \ref{subsec:free_QFT}, $\ket{0}$, are really the in state $\ket{0_-}$. Henceforth, I will drop the subscript `-' on the vacuum state.  

The free in-in generating functional is given by\footnote{Note that we can simplify Eq. (\ref{free_in-in_generating_functional}) further since $J_x^+ \Delta^-_{xx'}J_{x'}^-=J_x^- \Delta^+_{xx'}J_{x'}^+$.}
\begin{equation}
    \label{free_in-in_generating_functional}
    Z^{(0)}[J^+,J^-]=e^{-\frac{1}{2}\lb J_x^+ i\Delta^F_{xx'} J_{x'}^+-J_x^+ \Delta^-_{xx'}J_{x'}^--J_x^- \Delta^+_{xx'}J_{x'}^++J_x^-i\Delta^D_{xx'}J_{x'}^-\rb}
\end{equation}
such that all the free field 2-point functions given in Table \ref{tab:2-pt_functions} can be computed using the in-in formalism via Eq. (\ref{in-in_scalar_correlators}). To include quartic self-interactions, we write the generating functional as
\begin{equation}
    \label{interacting_generating_functional}
    Z[J^+,J^-]=e^{-i\frac{\lambda}{4} d^4x \sqrt{-g}\lb\frac{\delta^4}{\delta J^+(x)^4}-\frac{\delta^4}{\delta J^-(x)^4}\rb}Z^{(0)}[J^+,J^-].
\end{equation}
Applying Eq. (\ref{in-in_scalar_correlators}) to (\ref{interacting_generating_functional}), we can compute scalar correlators for this self-interacting theory.

\subsection{Two-point QFT correlation functions to $\mathcal{O}(\lambda)$}
\label{subsec:2-pt_function_one-loop}

The quantity of most interest for this work is the time-ordered 2-point function. Eq. (\ref{free_feynman_propagator}) gives us this quantity for free fields; now, we will add corrections from interactions in a perturbative manner. By expanding Eq. (\ref{interacting_generating_functional}) to leading order in small coupling $\lambda$, the generating functional to $\mathcal{O}(\lambda)$ is given by
\begin{equation}
    \label{O(lambda)_generating_functional}
    Z[J^+,J^-]=\lb1-i\frac{\lambda}{4}\int d^4x \sqrt{-g}\lb\frac{\delta^4}{\delta J^+(x)^4}-\frac{\delta^4}{\delta^4J^-(x)}\rb\rb Z^{(0)}[J^+,J^-].
\end{equation}
Using Eq. (\ref{in-in_scalar_correlators}), the time-ordered 2-point function to $\mathcal{O}(\lambda)$ is
\begin{equation}
    \label{one-loop_UVdiv_2-pt_function}
    \begin{split}
    \bra{0}T\hat{\phi}(x_1)&\hat{\phi}(x_2)\ket{0}=i\Delta^F(x_1,x_2)\\&+3i\lambda \expval{\hat{\phi}^2}\int d^4z a(t_z)^3\lsb i\Delta^F(z,x_1)i\Delta^F(z,x_2)-\Delta^+(z,x_1)\Delta^+(z,x_2)\rsb\\&+\mathcal{O}(\lambda^2).
    \end{split}
\end{equation}
The first line is just the free Feynman propagator (\ref{free_feynman_propagator}) while the second line gives the contribution to $\mathcal{O}(\lambda)$, which is yet to be computed, and we have used the fact that $i\Delta^F(z,z)=i\Delta^D(z,z)=\expval{\hat{\phi}^2}$. Note that for the time-ordered correlation functions, the Feynman propagators are sourced by $J^+$ while the Wightman functions are sourced by $J^-$.

The $\mathcal{O}(\lambda)$ contribution to the 2-point function can be computed in a similar way to the standard procedure in Minkowski, by making a correction to the mass. Applying the operator $\Box_{dS}+m^2$ to the 2-point function (\ref{one-loop_UVdiv_2-pt_function}), we find that at one-loop order it obeys the equation
\begin{equation}
    \label{2-pt_func_eom_O(lambda)}
    \lb\Box_{dS}+m_B^2+3\lambda\expval{\hat{\phi}^2}\rb\bra{0}T\phi(x_1)\phi(x_2)\ket{0}=-\frac{i}{a(t)^3}\delta^{(4)}(x_1-x_2),
\end{equation}
where the bare mass is $m_B^2=m_{0,B}^2+12\xi_B H^2$. We infer that the 2-point function 
to $\mathcal{O}(\lambda)$ can be computed simply by replacing the mass $m$ in the propagator by the Hubble-rate dependent effective mass
\begin{equation}
m_{\rm eff}^2(H)=m_B^2+3\lambda\expval{\hat{\phi}^2},
\end{equation}
i.e.,
\begin{equation}
    \label{O(lambda)_2-pt_function}
    \bra{0}T\hat{\phi}(x_1)\hat{\phi}(x_2)\ket{0}=i\Delta^F(x_1,x_2)\eval_{m^2=m_{\rm eff}^2(H)}.
\end{equation}
It is important to note that the effective mass $m_{\rm eff}^2(H)$ is finite, but both the bare mass $m_B^2$ and the field variance  $\expval{\hat{\phi}^2}$ are ultraviolet divergent. Therefore, the calculation requires regularisation and renormalisation. 
In order to make our effective theory directly applicable to particle physics theories, we want to use dimensional regularisation and the $\overline{\rm MS}$ renormalisation scheme, which is the convention in particle physics.

In dimensional regularisation, one takes the number of spacetime dimensions to be $d=4-\epsilon$. When one then takes the limit $d\rightarrow 4$, the UV divergence appears as a $1/\epsilon$ pole. For our purposes, we need the full expression for the field variance, including finite terms, and we are not aware of such a calculation in the literature. Therefore we present it in full detail in Appendix~\ref{app:dim_reg_de_Sitter}. The result is
\begin{equation}
    \label{field_variance_dim_reg_deSitter_ana}
    \expval{\hat{\phi}^2}=\frac{2H^2-m_B^2}{16\pi^2}\qty[\frac{2}{\epsilon}+\ln\frac{4\pi\mu^2}{a(t)^2H^2}-\gamma_E+1-\psi^{(0)}\qty(\frac{3}{2}-\nu_B)-\psi^{(0)}\qty(\frac{3}{2}+\nu_B)],
\end{equation}
where $\nu_B=\sqrt{\frac{9}{4}-\frac{m^2_B}{H^2}}$, $\psi^{(0)}(z)$ is the polygamma function, $\gamma_E$ is the Euler-Mascheroni constant, and $\mu$ is an arbitrary energy scale.

The next step is renormalisation, which involves absorbing the divergence into the mass parameter. In the $\overline{\text{MS}}$ scheme of dimensional regularisation, the renormalised mass is given by
\begin{equation}
\label{renormalised_mass_dim_reg_deSitter}
    \begin{split}
        m_R^2&=m_{0,R}^2+12\xi_RH^2\\&=m_B^2+\frac{3\lambda\qty(2H^2-m_B^2)}{16\pi^2}\qty(\frac{2}{\epsilon}-\gamma_E+\ln(\frac{4\pi\mu^2}{M^2}))+\mathcal{O}\qty(\lambda^2),
    \end{split}
\end{equation}
where $M$ is the renormalisation scale. Explicitly, we must renormalise both the scalar mass and the non-minimal coupling respectively as
\begin{subequations}
    \begin{align}
    \label{renormalised_mass_0}
        m_{0,R}^2&=m_{0,B}^2-\frac{3\lambda m_{0,B}^2}{16\pi^2}\qty(\frac{2}{\epsilon}-\gamma_E+\ln(\frac{4\pi\mu^2}{M^2}))+\mathcal{O}\qty(\lambda^2),\\
    \label{renormalised_non-minimal_coupling}
        \xi_R&=\xi_B+\frac{3\lambda\qty(\frac{1}{6}-\xi_B)}{16\pi^2}\qty(\frac{2}{\epsilon}-\gamma_E+\ln(\frac{4\pi\mu^2}{M^2}))+\mathcal{O}\qty(\lambda^2).
    \end{align}
\end{subequations}
We see that it is crucial to include the non-minimal coupling term for the renormalisation counterterms to be independent of $H$. Finally, we can now express the effective mass in Eq.~(\ref{O(lambda)_2-pt_function})
in terms of the $\overline{\rm MS}$ renormalied mass $m_R^2$ as
\begin{equation}
    \label{renormalised_mass_correction}
        m_{\rm eff}^2(H)= m_R^2+
\frac{3\lambda(2H^2-m_R^2)}{16\pi^2}
\qty[1-\psi^{(0)}\qty(\frac{3}{2}-\nu_R)
-\psi^{(0)}\qty(\frac{3}{2}+\nu_R)+\ln\frac{M^2}{a(t)^2H^2}].
\end{equation}
Note that the explicit dependence on the renormalisation scale $M$ cancels the implicit dependence through $m_R^2$, and therefore the effective mass $m^2_{\rm eff}(H)$ is renormalisation scale independent, as it must be.

For comparison with the stochastic approach, we will be interested in the long-distance behaviour of the 2-point function (\ref{O(lambda)_2-pt_function}). Focussing on spacelike separations, the leading term in the asymptotic expansion about long distances is
\begin{equation}
    \label{O(lambda)_2-pt_function_spacelike_long-distance}
    \begin{split}
    \bra{0}\hat{\phi}(t,\mathbf{0})\hat{\phi}(t,\mathbf{x})\ket{0}=&\frac{H^2}{16\pi^2}\frac{\Gamma\qty(3/2-\nu_R)\Gamma\qty(2\nu_R)4^{3/2-\nu_R}}{\Gamma\qty(1/2+\nu_R)}\Bigg[1+\frac{3 \lambda \qty(2H^2-m_R^2)}{32\pi^2\nu_RH^2}\\&\times\qty(\ln4+\psi^{(0)}\qty(3/2-\nu_R)-2\psi^{(0)}\qty(2\nu_R)+\psi^{(0)}\qty(1/2+\nu_R))\\&\times\qty(1-\psi^{(0)}\qty(3/2-\nu_R)-\psi^{(0)}\qty(3/2+\nu_R)+\ln\qty(\frac{M^2}{a(t)^2H^2}))\Bigg]\\&\times\abs{Ha(t)\mathbf{x}}^{-\frac{2\Lambda_1^{(QFT)}}{H}}+\mathcal{O}(\lambda^2),
    \end{split}
\end{equation}
where the exponent\footnote{We introduce the notation $\Lambda_1^{(QFT)}$ as a precursor to that used for the spectral expansion method in the stochastic theories.} is
\begin{equation}
    \label{QFT_exponent}
    \begin{split}
    \Lambda_1^{(QFT)}=&\qty(\frac{3}{2}-\nu_R)H
    \\&+\frac{3\lambda (2H^2-m_R^2)}{32\pi^2\nu_RH}\Bigg[1-\psi^{(0)}\qty(\frac{3}{2}-\nu_R)-\psi^{(0)}\qty(\frac{3}{2}+\nu_R)+\ln\frac{M^2}{a(t)^2H^2}\Bigg]+\mathcal{O}\qty(\lambda^2).
    \end{split}
\end{equation}
In principle, we could extend the calculation to higher orders in $\lambda$, albeit with increasing levels of complexity. However, there is a problem that stems from the IR limit hidden amongst our results. To see this, we expand the 2-point function (\ref{O(lambda)_2-pt_function_spacelike_long-distance}) to leading order in small-mass $m^2\ll H^2$ to give
\begin{equation}
    \label{O(lambda)_2-point_function_light_fields}
    \bra{0}\hat{\phi}(t,\mathbf{0})\hat{\phi}(t,\mathbf{x})\ket{0}=\lb\frac{3H^4}{8\pi^2m^2_R}-\frac{27\lambda H^8}{64\pi^4m^6_R}\rb\abs{Ha(t)\mathbf{x}}^{-\frac{2m^2_R}{3H^2}-\frac{3\lambda H^2}{4\pi^2m_R^2}}.
\end{equation}
One can see that both corrections to the amplitude and exponent are of relative order $\frac{\lambda H^4}{m_R^4}$ in this small-mass expansion. In order for the sum to converge at higher orders in $\lambda$, we require $\frac{\lambda H^4}{m_R^4}\ll1$. This is not \textit{a priori} true since our perturbation theory takes $\lambda$ to be the small parameter about which we expand. Thus, perturbative QFT is limited to the following region in the parameter space: $\lambda\ll1$ and $\lambda\ll m^4/H^4$. Therefore, to leading order in $\lambda H^4/m^4$, the long-distance behaviour of the spacelike field correlator is\footnote{Note that it was this 2-point function that was used in Ref. \cite{Cable:2022}. A key extension in this work is that we now consider Eq. (\ref{O(lambda)_2-pt_function_spacelike_long-distance}) when we match the perturbative QFT and stochastic 2-point functions. This introduces the $M$-dependence into the stochastic theory, which is an important UV effect even at long-distances.}
\begin{equation}
    \label{phi-phi_O(lambdaH4m4}
        \begin{split}
        \bra{0}\hat{\phi}(t,\mathbf{0})\hat{\phi}(t,\mathbf{x})\ket{0}=&\lsb\frac{H^2}{16\pi^2}\frac{\Gamma\qty(3/2-\nu_R)\Gamma\qty(2\nu_R)4^{3/2-\nu_R}}{\Gamma\qty(1/2+\nu_R)}-\frac{27\lambda H^8}{64\pi^4m^6_R}+\mathcal{O}\lb\frac{\lambda H^6}{m^4_R}\rb\rsb\\&\times\abs{Ha(t)\mathbf{x}}^{-(3-2\nu_R)-\frac{3\lambda H^2}{4\pi^2m_R^2}+\mathcal{O}(\lambda)},
        \end{split}
\end{equation}
This is as far as perturbative QFT will take us for 2-point correlation functions. In order to go beyond this, we must employ alternative methods, such as the stochastic effective theory of scalar fields in de Sitter. 

\subsection{Four-point QFT correlation functions to $\mathcal{O}(\lambda)$}
\label{subsec:4-pt_functions}

To round out our discussion of perturbative QFT, we will briefly consider the 4-pt functions. For this paper, we will largely consider them as a tool for computing stochastic parameters and so won't go into a huge amount of detail. However, it is important to recognise that these objects are computationally challenging and our work has raised some questions surrounding this, which we will touch on at the end of this section.

Using the Schwinger-Keldysh formalism outlined in Sec. \ref{subsec:in-in_formalism}, we can combine Eq. (\ref{in-in_scalar_correlators}) with (\ref{free_in-in_generating_functional}) and (\ref{interacting_generating_functional}) to obtain the time-ordered 4-pt scalar correlation function to $\mathcal{O}(\lambda)$ as
\begin{equation}
    \label{4-pt_function_Scwhinger-Keldysh}
    \begin{split}
        \bra{0}T\hat{\phi}(x_1)&\hat{\phi}(x_2)\hat{\phi}(x_3)\hat{\phi}(x_4)\ket{0}\\=&i\Delta^F(x_1,x_2)i\Delta^F(x_3,x_4)+[\text{2 more permutations}]
        \\&
        \begin{split}
        +3i\lambda\expval{\hat{\phi}^2}i\Delta^F(x_3,x_4)\int d^4z a(t_z)^3[ &i\Delta^F(z,x_1)i\Delta^F(z,x_2)\\&-\Delta^+(z,x_1)\Delta^+(z,x_2)]
        \end{split}
        \\&+[\text{5 more permutations}]
        \\&
        \begin{split}
        -6i\lambda\int d^4z a(t_z)^3[& i\Delta^F(z,x_1)i\Delta^F(z,x_2)i\Delta^F(z,x_3)i\Delta^F(z,x_4)\\&-\Delta^+(z,x_1)\Delta^+(z,x_2)\Delta^+(z,x_3)\Delta^+(z,x_4)].
        \end{split}
    \end{split}
\end{equation}
The first line after the equality sign is the free part, composed of a combination of Feynman propagators. The next three lines are a similar combination but this time occur at $\mathcal{O}(\lambda)$, thus containing the $\mathcal{O}(\lambda)$ piece of the 2-pt function (\ref{one-loop_UVdiv_2-pt_function}), multiplied by the free Feynman propagator. The final lines indicate the new contribution to the 4-pt function that first appears at $\mathcal{O}(\lambda)$. These terms make up the \textit{connected 4-point function}, which we will focus on here. Explicitly, this is related to the 2-point functions by
\begin{equation}
    \label{connected_4-pt_func_QFT}
    \begin{split}
    \bra{0}T\hat{\phi}(x_1)\hat{\phi}(x_2)\hat{\phi}(x_3)\hat{\phi}(x_4)\ket{0}_C=&\bra{0}T\hat{\phi}(x_1)\hat{\phi}(x_2)\hat{\phi}(x_3)\hat{\phi}(x_4)\ket{0}\\&-\bra{0}T\hat{\phi}(x_1)\hat{\phi}(x_2)\ket{0}\bra{0}T\hat{\phi}(x_3)\hat{\phi}(x_4)\ket{0}\\&-\bra{0}T\hat{\phi}(x_1)\hat{\phi}(x_3)\ket{0}\bra{0}T\hat{\phi}(x_2)\hat{\phi}(x_4)\ket{0}\\&-\bra{0}T\hat{\phi}(x_1)\hat{\phi}(x_4)\ket{0}\bra{0}T\hat{\phi}(x_2)\hat{\phi}(x_3)\ket{0},
    \end{split}
\end{equation}
where the subscript `\textit{C}' stands for `connected'. To compute this quantity, we must perform the $z$-integral in Eq. (\ref{4-pt_function_Scwhinger-Keldysh}). Since the integrand is composed of a series of hypergeometric functions, doing an analytic calculation is extremely difficult. Moreover, attempts at a numerical computation have proved fruitless due to the existence of poles in the integrand. Unfortunately, unlike its 2-pt counterpart, the 4-pt integral cannot be solved by a mass redefinition. We can make some progress by moving from position to momentum $\mathbf{k}$-space, as outlined in Ref. \cite{Serreau:2013,Serreau:2014,Nacir:2019}, where computations simplify and the pole structure is no longer a problem. For this purpose, we will focus on equal-time 4-pt functions. The Fourier transform goes as
\begin{equation}
    \label{fourier_transform}
    G^{(4)}_C(\eta,\{\mathbf{x}_i\})=\prod_{i=1}^4\lsb\int\dbar^3\mathbf{k}_ie^{-i\mathbf{k_i}\cdot\mathbf{x}_i}\rsb \Tilde{G}^{(4)}_C(\eta,\{\mathbf{k}_i\})\deltabar^{(3)}\lb\sum_{i=1}^4\mathbf{k}_i\rb,
\end{equation}
where $\deltabar^{(3)}(\mathbf{k})=(2\pi)^3\delta^{(3)}(\mathbf{k})$ and we use the shorthand notation for the connected 4-pt function $G^{(4)}_C(\eta,\{\mathbf{x}_i\})$, with $\{\mathbf{x}_i\}=(\mathbf{x}_1,\mathbf{x}_2,\mathbf{x}_3,\mathbf{x}_4)$, and the `tilde' indicates equivalent quantities in $\mathbf{k}$-space. Note that we will use conformal time in the following calculations, as defined in Eq. (\ref{conformal-->physical}). The equal-time connected 4-pt function in $\mathbf{k}$-space is given by
\begin{equation}
    \label{4-pt_func_k-space_general}
    \begin{split}
    \Tilde{G}^{(4)}_C(\eta,\{\mathbf{k}_i\})=&-6i\lambda\int^0_{-\infty} d\eta_z \frac{1}{(H\eta_z)^4}\\&
    \begin{split}
    \times\Bigg(&i\Tilde{\Delta}^F(\eta_z,\eta,\mathbf{k}_1)i\Tilde{\Delta}^F(\eta_z,\eta,\mathbf{k}_2)i\Tilde{\Delta}^F(\eta_z,\eta,\mathbf{k}_3)i\Tilde{\Delta}^F(\eta_z,\eta,\mathbf{k}_4)\\&-\Tilde{\Delta}^+(\eta_z,\eta,\mathbf{k}_1)\Tilde{\Delta}^+(\eta_z,\eta,\mathbf{k}_2) \Tilde{\Delta}^+(\eta_z,\eta,\mathbf{k}_3) \Tilde{\Delta}^+(\eta_z,\eta,\mathbf{k}_4)\Bigg), 
    \end{split}
    \end{split}
\end{equation}
where the Wightman function in $\mathbf{k}$-space is given by Eq. (\ref{k-space_2-pt_func}) and the Feynman propagator is found by using its definition in Table \ref{tab:2-pt_functions}. Using the time-ordering, one can simplify the integral such that
\begin{equation}
    \label{4-pt_func_k-space_Wightman}
\Tilde{G}^{(4)}_C(\eta,\{\mathbf{k}_i\})= \int _{-\infty}^\eta F(\eta,\eta_z,\{\mathbf{k}_i\}),
\end{equation}
where
\begin{equation}
    \begin{split}
     F(\eta,\eta_z,\{\mathbf{k}_i\})=&-6i\lambda \frac{1}{(H\eta_z)^4}\\&
    \begin{split}
    \times\Bigg(&\Tilde{\Delta}^-(\eta_z,\eta,\mathbf{k}_1)\Tilde{\Delta}^-(\eta_z,\eta,\mathbf{k}_2)\Tilde{\Delta}^-(\eta_z,\eta,\mathbf{k}_3)\Tilde{\Delta}^-(\eta_z,\eta,\mathbf{k}_4)\\&-\Tilde{\Delta}^+(\eta_z,\eta,\mathbf{k}_1)\Tilde{\Delta}^+(\eta_z,\eta,\mathbf{k}_2) \Tilde{\Delta}^+(\eta_z,\eta,\mathbf{k}_3) \Tilde{\Delta}^+(\eta_z,\eta,\mathbf{k}_4)\Bigg), 
    \end{split}
    \end{split}
\end{equation}
and $\Tilde{\Delta}^-(\eta_z,\eta,\mathbf{k})$ is the complex conjugate of $\Tilde{\Delta}^+(\eta_z,\eta,\mathbf{k})$.

This integral is hard to solve in general. Analytic solutions are difficult because the integrand is a product of Hankel functions whilst the oscillatory behaviour of the integrand in the limit $\eta_z\rightarrow-\infty$ make numerical computations challenging. However, for the purposes of this paper, we are only interested in the long-distance behaviour of the correlator. Assuming that that all four momenta  $k_i$ $\forall i\in\{1,2,3,4\}$ are of the same order of magnitude, which we denote by $k$, we can therefore assume that $-k\eta\ll 1$.
We then separate the integral at some intermediate time $\eta_0=-\Lambda/k<\eta$, 
where $\Lambda\ll1$ such that $-k\eta_0\ll1$. Then, the 4-point function can be written as
\begin{equation}
    \label{4-pt_func_split}
    \Tilde{G}^{(4)}_C(\eta,\{\mathbf{k}_i\})=\int^{\eta_0}_{-\infty}d\eta_z F(\eta,\eta_z,\{\mathbf{k}_i\})+\int^\eta_{\eta_0}d\eta_z F(\eta,\eta_z,\{\mathbf{k}_i\}),
\end{equation}
Considering the $k\eta\ll-1$ limit of the Wightman functions,
\begin{equation}
    \label{Wightman_k-space_Hankels_kll1}
    \begin{split}
    \Tilde{\Delta}^{\pm}(\eta_z,\eta,\mathbf{k})\simeq & \frac{\pi}{4Ha(\eta)^{3/2}a(\eta_z)^{3/2}}\mathcal{H}_{\nu_R}^{(1)/(2)}(-k\eta_z) \times \\&
\Bigg(\frac{2^{-\nu_R}}{\nu_R\Gamma(\nu_R)}(-k\eta)^{\nu_R}\pm i\frac{2^{\nu_R}\Gamma(\nu_R)}{\pi}(-k\eta_z)^{-\nu_R}\Bigg),
    \end{split}
\end{equation}
one observes that the first term in Eq.~(\ref{4-pt_func_split}) is proportional to the power law  $k^{-3-4\nu_R}$. We show this more explicitly in Appendix \ref{app:IR_lim_4-pt_func}.

For the second term, we can make use of the approximation $-k\eta_z\ll1$ to write the Wightman functions as 
\begin{equation}
    \label{k_Wightman_IR_lim}
    \begin{split}
    \Tilde{\Delta}^\pm(\eta_z,\eta,\mathbf{k})\simeq \frac{\pi}{4Ha(\eta)^{3/2}a(\eta_z)^{3/2}}\Bigg[&\frac{4^{\nu_R}\Gamma(\nu_R)^2}{\pi^2}\lb\eta_z\eta\rb^{-\nu_R}k^{-2\nu_R}\\&\pm i\frac{1}{\pi\nu_R}\lb\lb\frac{\eta}{\eta_z}\rb^{\nu_R}-\lb\frac{\eta_z}{\eta}\rb^{\nu_R}\rb\Bigg].
    \end{split}
\end{equation}
Substituting Eq. (\ref{k_Wightman_IR_lim}) into the second term of  (\ref{4-pt_func_split}), one can compute the integral to give
\begin{equation}
    \label{k-space_4-pt_func}
    \begin{split}
    \Tilde{G}_C^{(4)}(\eta,\{\mathbf{k}_i\}&)\simeq\mathcal{O}(k^{-3-4\nu_R})-\frac{3\lambda}{4H^5}\lb\frac{4^{\nu_R}\Gamma(\nu_R)^2}{2\pi}\rb^3\frac{1}{(3-4\nu_R)(3-2\nu_R)}(-H\eta)^{9-6\nu_R}
    \\&\times\lsb\lb\frac{k_1k_2k_3}{H^3}\rb^{-2\nu_R}+\lb\frac{k_1k_3k_4}{H^3}\rb^{-2\nu_R}+\lb\frac{k_2k_3k_4}{H^3}\rb^{-2\nu_R}+\lb\frac{k_1k_2k_4}{H^3}\rb^{-2\nu_R}\rsb,
    \end{split}
\end{equation}
where the first term includes contributions from the the first term of Eq.~(\ref{4-pt_func_split}) and from  the lower limit of the second integral, which are both of the same order $\mathcal{O}(k^{-3-4\nu_R})$. This is actually the leading contribution in the IR limit for light fields, over the second term in Eq.~(\ref{k-space_4-pt_func}), which is of order $\mathcal{O}(k^{-6\nu_R})$. All other contributions are subdominant to $k^{-6\nu_R}$. For a deeper discussion of this, see Appendix \ref{app:IR_lim_4-pt_func}.

It is challenging to get analytic results for the leading term $\mathcal{O}(k^{-3-4\nu_R})$, especially in coordinate space, because it will depend on all 4 $k_i$s simultaneously and thus the $\delta$-function arising in the Fourier transform (\ref{fourier_transform}) will result in a mixing of momenta. On the other hand, the $\mathcal{O}(k^{-6\nu_R})$ contribution will deal with the $\delta$-function trivially because each term only ever depends on 3 of the 4 momenta. For this work, it is sufficient to have an analytic expression for one of the leading IR terms so that we can do a comparison with the stochastic approach. However, this does leave the door open for more careful analysis of the 4-pt functions.

To convert Eq. (\ref{k-space_4-pt_func}) to coordinate space, we can use the Fourier transform (\ref{fourier_transform}), using the result \cite{Nacir:2019}
\begin{equation}
    \label{Nacir_identity}
    \int \dbar^3\mathbf{k}e^{-i\mathbf{k}\cdot\mathbf{x}}k^{w-3}=\frac{1}{(2\pi)^3}\frac{2^{3-2\nu_R}\pi^{3/2}\Gamma\lb\frac{5}{2}-\nu_R\rb}{\lb\frac{3}{2}-\nu_R\rb\Gamma(\nu_R)}x^{-w},
\end{equation}
to obtain the equal-time connected 4-pt function in coordinate space
\begin{equation}
    \label{spacelike_quantum_4-pt_func}
    \bra{0}\hat{\phi}(t,\mathbf{x}_1)\hat{\phi}(t,\mathbf{x}_2)\hat{\phi}(t,\mathbf{x}_3)\hat{\phi}(t,\mathbf{x}_4)\ket{0}=...+\frac{3\lambda H^4}{\pi^{15/2}}\frac{\Gamma(\nu_R)^3\Gamma\lb\frac{5}{2}-\nu_R\rb^3}{\lb4\nu_R-3\rb\lb3-2\nu_R\rb^4}\abs{Ha(t)\mathbf{x}}^{-9+6\nu_R},
\end{equation}
where $\abs{\mathbf{x}}=\abs{\mathbf{x}_i-\mathbf{x}_j}$ $\forall i\ne j$, $i,j\in\{1,2,3,4\}$ and the `$...+$' indicates the other leading IR contribution $\mathcal{O}(\abs{Ha(t)\mathbf{x}}^{-6+4\nu_R})$. Note that, for light fields, this contribution is given by 
\begin{equation}
    \label{spacelike_quantum_4-pt_func_light_fields}
    \bra{0}\hat{\phi}(t,\mathbf{x}_1)\hat{\phi}(t,\mathbf{x}_2)\hat{\phi}(t,\mathbf{x}_3)\hat{\phi}(t,\mathbf{x}_4)\ket{0}\eval_{m^2\ll H^2}=...+\frac{81\lambda H^{12}}{128\pi^6m_R^8}\abs{Ha(t)\mathbf{x}}^{-\frac{2m^2_R}{H^2}}.
\end{equation}

\section{Second-order stochastic effective theory}
\label{sec:second-order_theory}

We will now consider the second-order stochastic effective theory, which was introduced in Ref. \cite{Cable:2021,Cable:2022}. We again consider a scalar field with a mass $m_S$ and quartic self-interaction $\lambda_S$, where the subscript `S' stands for `stochastic', to differentiate from the QFT quantities introduced above. The stochastic equations are then given by
\begin{equation}
    \label{2d_stochastic_eq}
    \begin{pmatrix}\Dot{\phi}\\\Dot{\pi}\end{pmatrix}=\begin{pmatrix}\pi\\-3H\pi-m^2_S\phi-\lambda_S\phi^3\end{pmatrix}+\begin{pmatrix}\xi_{\phi}\\\xi_{\pi}\end{pmatrix}
\end{equation}
with a white noise contribution
\begin{equation}
    \label{white_noise_general}
    \expval{\xi_i(t)\xi_j(t')}=\sigma_{ij}^2\delta(t-t').
\end{equation}
The noise amplitudes $\sigma_{ij}^2$ are left unspecified for the time being, other than the fact that they do not depend on the spacetime coordinates and that they are symmetric, preserving the reality of the noise. The form of the stochastic parameters $m_S$, $\lambda_S$ and $\sigma_{ij}^2$ will be determined by comparing stochastic correlators with their perturbative QFT counterparts. We will choose these quantities to be the same, promoting our stochastic theory from something general to an effective theory of QFT.

\subsection{The second-order stochastic correlators}
\label{subsec:stochastic_correlators}

\subsubsection{The spectral expansion}
\label{subsubsec:spectral_expansion}

The time-evolution of the one-point probability distribution function (1PDF) $P(\phi,\pi;t)$ associated with the stochastic equations (\ref{2d_stochastic_eq}) is described by the Fokker-Planck equation
\begin{equation}
    \label{phi-pi_fokker-planck_eq}
    \begin{split}
        \partial_t P(\phi,\pi;t)=&\Bigg[3H-\pi\partial_\phi+(3H\pi+V'(\phi))\partial_\pi+\frac{1}{2}\sigma_{\phi\phi}^2\partial_\phi^2+\sigma_{\phi\pi}^2\partial_\phi\partial_\pi+\frac{1}{2}\sigma_{\pi\pi}^2\partial_\pi^2\Bigg]P(\phi,\pi;t)\\
        =&\mathcal{L}_{FP}P(\phi,\pi;t),
    \end{split}
\end{equation}
where $\mathcal{L}_{FP}$ is the Fokker-Planck operator. For a space of functions $\{f|(f,f)<\infty\}$ with the inner product
\begin{equation}
    \label{scalar_product}
    (f,g)=\int_{-\infty}^{\infty} d\phi\int_{-\infty}^{\infty} d\pi f(\phi,\pi)g(\phi,\pi),
\end{equation}
we define the adjoint of the Fokker-Planck operator, $\mathcal{L}_{FP}^*$, as
\begin{equation}
    \label{adjoint_fp_op_definition}
    \qty(\mathcal{L}_{FP}f,g)=\qty(f,\mathcal{L}_{FP}^*g).
\end{equation}
Note that all integrals over $\phi$ and $\pi$ have the limits $(-\infty,\infty)$ unless otherwise stated. Explicitly,
\begin{equation}
    \label{adjoint_fokker-planck_operator}
    \begin{split}
        \mathcal{L}_{FP}^*=&\pi\partial_{\phi}-\qty(3H\pi+V'(\phi))\partial_{\pi}+\frac{1}{2}\sigma_{\phi\phi}^2\partial_{\phi}^2+\sigma_{\phi\pi}^2\partial_{\phi}\partial_{\pi}+\frac{1}{2}\sigma_{\pi\pi}^2\partial_{\pi}^2.
    \end{split}    
\end{equation}
The 1PDF can be written as a spectral expansion
\begin{equation}
    \label{spectral_expansion_1PDF}
    P(\phi,\pi;t)=\Psi_{0}^*(\phi,\pi)\sum_{N=0}^{\infty}c_N\Psi_{N}(\phi,\pi)e^{-\Lambda_{N}t},
\end{equation}
where $\Lambda_{N}$ and $\Psi_{N}^{(*)}(\phi,\pi)$ are the respective eigenvalues and (adjoint) eigenstates to the (adjoint) Fokker-Planck operator
\begin{subequations}
    \label{FP_eigenequations}
    \begin{align}
        \mathcal{L}_{FP}\Psi_{N}(\phi,\pi)&=-\Lambda_{N}\Psi_{N}(\phi,\pi),\\
        \mathcal{L}_{FP}^*\Psi_{N}^*(\phi,\pi)&=-\Lambda_{N}\Psi^*_{N}(\phi,\pi),
    \end{align}
\end{subequations}
and $c_N$ are coefficients. We consider eigenstates that obey the biorthogonality and completeness relations
\begin{subequations}
    \label{biorthogonal&completeness_relations}
    \begin{align}
        \qty(\Psi_{N}^*,\Psi_{N'})&=\delta_{N'N},\\
        \sum_{N}\Psi_{N}^*(\phi,\pi)\Psi_{N}(\phi',\pi')&=\delta(\phi-\phi')\delta(\pi-\pi'),
    \end{align}
\end{subequations}
\noindent and there exists an equilibrium state $P_{eq}(\phi,\pi)=\Psi_{0}^*(\phi,\pi)\Psi_{0}(\phi,\pi)$ obeying $\partial_tP_{eq}(\phi,\pi)=0$. All sums of this form run from $N=0$ to $N=\infty$. Note that the  $\Psi_0^*(\phi,\pi)$ eigenstate, corresponding to $\Lambda_0=0$, is a constant, such that Eq. (\ref{spectral_expansion_1PDF}) obeys the Fokker-Planck equation (\ref{phi-pi_fokker-planck_eq}) \footnote{This is convenient when one considers the simplified case for free fields e.g. Eq. (\ref{free_eigenstates}). In this case, the non-adjoint and adjoint eigenstates only differ by a factor of a Gaussian.}. 

To obtain stochastic correlators, we introduce the transfer matrix $U(\phi_0,\phi,\pi_0,\pi;t-t_0)$ between $(\phi_0,\pi_0)=(\phi(t_0,\mathbf{x}),\pi(t_0,\mathbf{x}))$ and $(\phi,\pi)=(\phi(t,\mathbf{x}),\pi(t,\mathbf{x}))$, which is defined as the Green's function of the Fokker-Planck equation 
\begin{equation}
    \label{time-evolution_op_FP_equation}
    \partial_tU(\phi_0,\phi,\pi_0,\pi;t-t_0)=\mathcal{L}_{FP}U(\phi_0,\phi,\pi_0,\pi;t-t_0)
\end{equation}
for all values of $\phi_0$ and $\pi_0$. Then, the time-dependence of the 1PDF is given by
\begin{equation}
    \label{time-dependence_1PDF}
    P(\phi,\pi;t)=\int d\phi_0\int d\pi_0 P(\phi_0,\pi_0;t_0)U(\phi_0,\phi,\pi_0,\pi;t-t_0).
\end{equation}
From Eq. (\ref{spectral_expansion_1PDF}), making use of the relations (\ref{biorthogonal&completeness_relations}), we find that the transfer matrix can be written with the spectral expansion as
\begin{equation}
    \label{spectral_expansion_time-ev_operator}
    U(\phi_0,\phi,\pi_0,\pi;t-t_0)=\frac{\Psi_{0}^*(\phi,\pi)}{\Psi_{0}^*(\phi_0,\pi_0)}\sum_{N}\Psi_{N}^*(\phi_0,\pi_0)\Psi_{N}(\phi,\pi)e^{-\Lambda_{N}(t-t_0)}.
\end{equation}

\subsubsection{Two-point stochastic correlation functions}
\label{subsubsec:2-pt_2O_correlators}

We can write an equilibrium 2-point probability distribution function (2PDF) as
\begin{equation}
    \label{2PDF}
    \begin{split}
        P_2(\phi_0,\phi,\pi_0,\pi;t-t_0)&=P(\phi_0,\pi_0;t_0)U(\phi_0,\phi,\pi_0,\pi;t-t_0)\\
        &=\Psi_{0}^*(\phi,\pi)\Psi_{0}(\phi_0,\pi_0)\sum_{N}\Psi_{N}^*(\phi_0,\pi_0)\Psi_{N}(\phi,\pi)e^{-\Lambda_{N}(t-t_0)},
    \end{split}
\end{equation}
where we take $P(\phi_0,\pi_0;t_0)=P_{eq}(\phi_0,\pi_0)$. Then, the 2-point timelike (equal-space) stochastic correlator between some functions $f(\phi_0,\pi_0)$ and $g(\phi,\pi)$ is given by
\begin{equation}
    \label{2pt_timelike_stochastic_correlator_fg}
    \begin{split}
        \expval{f(\phi_0,\pi_0)g(\phi,\pi)}=&\int d\phi_0\int d\phi\int d\pi_0\int d\pi P_2(\phi_0,\phi,\pi_0,\pi;t-t_0)f(\phi_0,\pi_0)g(\phi,\pi)\\
        =&\sum_{N}f_{0N}g_{N0}e^{-\Lambda_{N}(t-t_0)},
    \end{split}
\end{equation}
where
\begin{equation}
    \label{fnl}
        f_{NN'}=\qty(\Psi_{N},f\Psi_{N'}^*).
\end{equation}
We can compute spacelike correlators by defining an equilibrium 3-point probability distribution function (3PDF), where we evolve both $(\phi_1,\pi_1)$ and $(\phi_2,\pi_2)$ to $(\phi_0,\pi_0)$ independently, as 
\begin{equation}
    \label{3PDF_spectral_expansion}
    \begin{split}
        P_3^{(S)}(\phi_0,\phi_1,\phi_2,\pi_0,\pi_1,\pi_2;&t_0,t_1,t_2)\\=&P(\phi_0,\pi_0;t_0)U(\phi_0,\phi_1,\pi_0,\pi_1;t_1-t_0)U(\phi_0,\phi_2,\pi_0,\pi_2;t_2-t_0)\\
        =&\frac{\Psi_{0}(\phi_0,\pi_0)\Psi^*_{0}(\phi_1,\pi_1)\Psi^*_{0}(\phi_2,\pi_2)}{\Psi^*_{0}(\phi_0,\pi_0)}\sum_{N}\Psi_{N}^*(\phi_0,\pi_0)\Psi_{N}(\phi_1,\pi_1)\\&\times\sum_{N'}\Psi_{N'}^*(\phi_0,\pi_0)\Psi_{N'}(\phi_2,\pi_2)e^{-\qty(\Lambda_{N}(t_1-t_0)+\Lambda_{N'}(t_2-t_0))},
    \end{split}
\end{equation}
where the superscript $(S)$ indicates that it is the 3PDF used to define spacelike correlators\footnote{We could similarly define a 3PDF for computing timelike correlators, where the evolution would be from $(\phi_0,\pi_0)$ to $(\phi_1,\pi_1)$ to $(\phi_2,\pi_2)$ i.e. chronologically along a line of constant spatial coordinate (assuming $t_0<t_1<t_2$).}. For more details, see Ref. \cite{Cable:2022}. To evaluate the spacelike (equal-time) stochastic correlators, we compute the 3-point function between two timelike separated points $t_1$ and $t_2$ and the time coordinate $t_r$, defined as
\begin{equation}
    \label{spacelike_time_coord}
    t_r=-\frac{1}{H}\ln\lb H\abs{\mathbf{x}_1-\mathbf{x}_2}\rb.
\end{equation}
The spacelike stochastic correlator between the functions $f(\phi(t,\mathbf{x}_1),\pi(t,\mathbf{x}_1))$ and $g(\phi(t,\mathbf{x}_2),\pi(t,\mathbf{x}_2))$ is given by integrating over $\phi_r$ and $\pi_r$ as
\begin{equation}
    \label{2pt_spacelike_stochastic_correlators_fg}
    \begin{split}
        \langle& f(\phi,\pi;t,\mathbf{x}_1)g(\phi,\pi;t,\mathbf{x}_2)\rangle\\=&\int d\phi_r\int d\phi_1\int d\phi_2\int d\pi_r\int d\pi_1\int d\pi_2 P_3(\phi_r,\phi_1,\phi_2,\pi_r,\pi_1,\pi_2;t_r,t_1,t_2)\\&\times f(\phi_1,\pi_1)g(\phi_2,\pi_2)
        \\=&\int d\phi_r\int d\pi_r \frac{\Psi_{0}(\phi_r,\pi_r)}{\Psi_{0}^*(\phi_r,\pi_r)}\sum_{NN'}\Psi_{N}^*(\phi_r,\pi_r)\Psi_{N'}^*(\phi_r,\pi_r)f_{N}g_{N'}\abs{Ha(t)(\mathbf{x}_1-\mathbf{x}_2)}^{-\frac{\Lambda_{N}+\Lambda_{N'}}{H}}.
    \end{split}
\end{equation}

\subsubsection{Four-point stochastic correlation functions}
\label{subsubsec:4-pt_2o_stochastic_correlator}

We can similarly compute 4-point functions via the spectral expansion. For the timelike 4-point functions, we write the equilibrium 4-point probability distribution function (4PDF) as
\begin{equation}
    \label{4PDF}
    \begin{split}
        P_4^{(T)}&(\phi_1,\phi_2,\phi_3,\phi_4,\pi_1,\pi_2,\pi_3,\pi_4;t_1,t_2,t_3,t_4)\\:=&P(\phi_1,\pi_1;t_1)U(\phi_1,\phi_2,\pi_1,\pi_2;t_2-t_1)U(\phi_2,\phi_3,\pi_2,\pi_3;t_3-t_2)U(\phi_3,\phi_4,\pi_3,\pi_4;t_4-t_3)\\
        =&\Psi_0(\phi_1,\pi_1)\Psi^*_0(\phi_4,\pi_4)\sum_{N}\Psi_N^*(\phi_1,\pi_1)\Psi_N(\phi_2,\pi_2)\sum_{N'}\Psi_{N'}^*(\phi_2,\pi_2)\Psi_{N'}(\phi_3,\pi_3)\\&\times\sum_{N''}\Psi_{N''}^*(\phi_3,\pi_3)\Psi_{N''}(\phi_4,\pi_4),
    \end{split}
\end{equation}
where the superscript $(T)$ indicates we are using this 4PDF to compute timelike correlators. Assuming that $t_1<t_2<t_3<t_4$, and that $P(\phi_1,\pi_1;t_1)=P_{eq}(\phi_1,\pi_1)$, the timelike 4-point correlation function is given by
\begin{equation}
    \label{4-pt_stochastic_function_general}
    \begin{split}
        \langle f_1(\phi_1,\pi_1)f_2&(\phi_2,\pi_2)f_3(\phi_3,\pi_3)f_4(\phi_4,\pi_4)\rangle\\=&\prod_{i=1}^4\int d\phi_i\int d\pi_i P_4^{(T)}(\phi_1,\phi_2,\phi_3,\phi_4,\pi_1,\pi_2,\pi_3,\pi_4;t_1,t_2,t_3,t_4)\\&\times f_1(\phi_1,\pi_1)f_2(\phi_2,\pi_2)f_3(\phi_3,\pi_3)f_4(\phi_4,\pi_4)\\
        =&\sum_{N''N'N}(f_1)_{0N}(f_2)_{NN'}(f_3)_{N'N''}(f_4)_{N''0}e^{-\Lambda_N(t_2-t_1)-\Lambda_{N'}(t_3-t_2)-\Lambda_{N''}(t_4-t_3)}.
    \end{split}
\end{equation}
In a similar computation to the 2-point function, we can compute the spacelike 4-point function. Now, we define the ``spacelike'' equilibrium 5-point probability distribution function (5PDF) as
\begin{equation}
    \label{5PDF}
    \begin{split}
        P_5^{(S)}&(\phi_0,\phi_1,\phi_2,\phi_3,\phi_4,\pi_1,\pi_2,\pi_3,\pi_4;t_0,t_1,t_2,t_3,t_4)\\:=&P(\phi_0,\pi_0;t_0)U(\phi_0,\phi_1,\pi_0,\pi_1;t_1-t_0)U(\phi_1,\phi_2,\pi_1,\pi_2;t_2-t_0)U(\phi_2,\phi_3,\pi_2,\pi_3;t_3-t_0)\\&\times U(\phi_3,\phi_4,\pi_3,\pi_4;t_4-t_0)\\
        =&\frac{\Psi_0(\phi_0,\pi_0)}{\Psi_0^*(\phi_0,\pi_0)^3}\prod_{i=1}^4\lsb\Psi_0^*(\phi_i,\pi_i)\sum_N\Psi_N^*(\phi_0,\pi_0)\Psi_N(\phi_i,\pi_i)\rsb.
    \end{split}
\end{equation}
Using the $t_r$ coordinate in Eq. (\ref{spacelike_time_coord}) and assuming $\abs{\mathbf{x}}=\abs{\mathbf{x}_i-\mathbf{x}_j}\text{ }\forall i\ne j$, the spacelike stochastic 4-point function between some functions $f_i(\phi,\pi)$, $i\in\{1,2,3,4\}$, is given by
\begin{equation}
    \label{spacelike_4-pt_correlator_f}
    \begin{split}
        \langle& f_1(\phi_1,\pi_1)f_2(\phi_2,\pi_2)f_3(\phi_3,\pi_3)f_4(\phi_4,\pi_4)\rangle\\
        =&\int d\phi_r\int d\pi_r\prod_{i=1}^4\int d\phi_i\int d\pi_i P_5^{(S)}(\phi_r,\phi_1,\phi_2,\phi_3,\phi_4,\pi_r,\pi_1,\pi_2,\pi_3,\pi_4;t_r,t_1,t_2,t_3,t_4)\\&\times f_1(\phi_1,\pi_1)f_2(\phi_2,\pi_2)f_3(\phi_3,\pi_3)f_4(\phi_4,\pi_4)\\
        =&\int d\phi_r\int d\pi_r\frac{\Psi_0(\phi_r,\pi_r)}{\Psi_0^*(\phi_r,\pi_r)^3}\prod_{i=1}^4\lsb\sum_N\Psi_N^*(\phi_r,\pi_r)(f_i)_{N0}\abs{Ha(t)\mathbf{x}}^{-\frac{\Lambda_N}{H}}\rsb.
    \end{split}
\end{equation}

\subsection{Comparison with perturbative QFT}
\label{subsec:comp_perturbative_QFT}

Now that we have developed the formalism for the stochastic theory, we will now promote it to an effective theory of the IR behaviour of scalar fields in de Sitter spacetime. To do this, we will compare the stochastic and QFT correlators and choose the stochastic parameters such that they match. This procedure was first outlined in Ref. \cite{Cable:2021,Cable:2022}.

\subsubsection{Free stochastic parameters}
\label{subsubsec:free_stochastic_parameters}

We will begin by considering free fields, which was first done in Ref. \cite{Cable:2021}. It will prove convenient to change our field variables from $(\phi,\pi)$ to $(q,p)$, with the transformation
\begin{equation}
    \label{(q,p)=A(phi,pi)}
    \begin{pmatrix}p\\q\end{pmatrix}=\frac{1}{\sqrt{1-\frac{\alpha_S}{\beta_S}}}\begin{pmatrix}1&\alpha_S H\\\frac{1}{\beta_S H}&1\end{pmatrix}\begin{pmatrix}\pi\\\phi\end{pmatrix},
\end{equation}
where $\alpha_S=\frac{3}{2}-\nu_S$ and $\beta_S=\frac{3}{2}+\nu_S$ with $\nu_S=\sqrt{\frac{9}{4}-\frac{m^2_S}{H^2}}$. All of the formalism introduced in the previous section can also be applied to $(q,p)$ variables. In particular the Fokker-Planck operators are given by
\begin{subequations}
    \label{free+int_FP_operators}
    \begin{align}
        \mathcal{L}_{FP}&=\mathcal{L}_{FP}^{(0)}+\lambda\mathcal{L}_{FP}^{(1)},\\
        \mathcal{L}_{FP}^*&=\mathcal{L}_{FP}^{(0)*}+\lambda\mathcal{L}_{FP}^{(1)*},
    \end{align}
\end{subequations}
where the free part is given by
\begin{subequations}
    \label{free_FP_operators}
    \begin{align}
        \mathcal{L}_{FP}^{(0)}&=\alpha H+\alpha H q\partial_q+\frac{1}{2}\sigma_{qq}^{2}\partial_q^2+\beta H +\beta H p\partial_p+\frac{1}{2}\sigma_{pp}^{2}\partial_p^2+\sigma_{qp}^{2}\partial_q\partial_p,\\
        \mathcal{L}_{FP}^{(0)*}&=-\alpha H q\partial_q+\frac{1}{2}\sigma_{qq}^{2}\partial_q^2-\beta H p\partial_p+\frac{1}{2}\sigma_{pp}^{2}\partial_p^2+\sigma_{qp}^{2}\partial_q\partial_p
    \end{align}
\end{subequations}
and the interacting part is given by
\begin{subequations}
    \label{interacting_FP_operators}
    \begin{align}
        \mathcal{L}_{FP}^{(1)}&=\frac{\lambda}{(1-\frac{\alpha}{\beta})^2}\qty(-\frac{1}{\beta H}p+q)^3\qty(\partial_p+\frac{1}{\beta H}\partial_q),\\
        \mathcal{L}_{FP}^{(1)*}&=-\mathcal{L}_{FP}^{(1)}.
    \end{align}
\end{subequations}
The $(q,p)$ noise amplitudes are written in terms of their $(\phi,\pi)$ counterparts as
\begin{subequations}
    \label{sigma_q,p-->sigma_phi,pi}
    \begin{align}
        \sigma_{qq}^2=&\frac{1}{1-\frac{\alpha}{\beta}}\qty(\frac{1}{\beta^2H^2}\sigma_{\pi\pi}^2+\frac{2}{\beta H}\sigma_{\phi\pi}^2+\sigma_{\phi\phi}^2),\\
        \sigma_{qp}^2=&\frac{1}{1-\frac{\alpha}{\beta}}\qty(\frac{1}{\beta H}\sigma_{\pi\pi}^2+\qty(1+\frac{\alpha}{\beta })\sigma_{\phi\pi}^2+\alpha H\sigma_{\phi\phi}^2),\\
        \sigma_{pp}^2=&\frac{1}{1-\frac{\alpha}{\beta}}\qty(\sigma_{\pi\pi}^2+2\alpha H\sigma_{\phi\pi}^2+\alpha^2H^2\sigma_{\phi\phi}^2).
    \end{align}
\end{subequations}
Following the work of Ref. \cite{Cable:2021}, we compute the stochastic free field correlator as
\begin{equation}
    \label{free_stochastic_field_correlator}
    \begin{split}
    \expval{\phi(t,\mathbf{0})\phi(t,\mathbf{x})}=&\frac{1}{1-\frac{\alpha_S}{\beta_S}}\Bigg[\frac{\sigma_{qq}^2}{2H\alpha_S}\abs{Ha(t)\mathbf{x}}^{-2\alpha_S}+\frac{\sigma_{pp}^2}{2H^3\beta_S^3}\abs{Ha(t)\mathbf{x}}^{-2\beta_S}\\&-\frac{2\sigma_{qp}^2}{3H^2\beta_S}\abs{Ha(t)\mathbf{x}}^{-3}\Bigg],
    \end{split}
\end{equation}
and match it to the free Feynman propagator (\ref{free_feynman_long-distance_leading}) to obtain an expression for the free stochastic parameters
\begin{subequations}
    \label{free_matched_stochastic_parameters}
\begin{align}
        \label{matched_mass}
        m_S^{(0)}&=m\\
        \label{matched_sigma_qq}
        \sigma_{qq}^{(0)2}&=\sigma_{Q,qq}^{2(0)}=\frac{H^3\alpha\nu}{4\pi^2\beta}\frac{\Gamma(2\nu)\Gamma(\frac{3}{2}-\nu)4^{\frac{3}{2}-\nu}}{\Gamma(\frac{1}{2}+\nu)},\\
        \label{matched_sigma_qp}
        \sigma_{qp}^{(0)2}&=\sigma_{Q,qp}^{2(0)}=0,\\
        \label{matched_sigma_pp}
        \sigma_{pp}^{(0)2}&=\sigma_{Q,pp}^{2(0)}=\begin{cases}\sigma_{pp}^{2(NLO)(0)}=\frac{H^5\beta^2\nu}{4\pi^2}\frac{\Gamma(-2\nu)\Gamma(\frac{3}{2}+\nu)4^{\frac{3}{2}+\nu}}{\Gamma(\frac{1}{2}-\nu)}\\0\end{cases}.
\end{align}
\end{subequations}
$m_S^{(0)}$ and $\sigma_{qq}^{2(0)}$ are matched such that the leading-order exponent and amplitude in the Feynman propagator are reproduced while $\sigma_{qp}^{2(0)}$ noise is chosen such that there is an analytic continuation from timelike to spacelike stochastic correlators, a behaviour prevalent in QFT. However, the choice of $\sigma_{pp}^2$ is arbitrary. In this paper, we focus on two possible choices: that the subleading term in the Feynman propagator is reproduced ($\sigma_{Q,pp}^2=\sigma_{pp}^{2(NLO)}$) or that the subleading term in the stochastic field 2-point function vanishes ($\sigma_{Q,pp}^2=0$). We will see later that this choice doesn't impact physical results. 

Since $\sigma_{qp}^{2(0)}=0$, the variables $p$ and $q$ separate and so we now use two indices $(r,s)\in\{0,\infty\}$, corresponding to $p$ and $q$ respectively, as opposed to just $N$. Thus, the free field eigenquations are given by
\begin{subequations}
    \label{FP_free_eigenequations}
    \begin{align}
        \mathcal{L}^{(0)}_{FP}\Psi_{rs}^{(0)}(q,p)&=-\Lambda_{rs}^{(0)}\Psi_{rs}^{(0)}(q,p),\\
        \mathcal{L}_{FP}^{(0)*}\Psi_{rs}^{(0)*}(q,p)&=-\Lambda_{rs}^{(0)}\Psi^{(0)*}_{rs}(q,p),
    \end{align}
\end{subequations}
where the $\Lambda_{rs}^{(0)}$ and $\Psi_{rs}^{(0)(*)}(q,p)$ are the free eigenvalues and (adjoint) eigenstates respectively. The eigenvalues of Eq. (\ref{FP_free_eigenequations}) are
\begin{equation}
    \label{free_eigenvalues}
    \Lambda_{rs}^{(0)}=\qty(s\alpha+r\beta)H
\end{equation}
while the normalised eigenstates can be written in terms of the Hermite polynomials $H_n(x)$ as
\begin{subequations}
    \label{free_eigenstates}
    \begin{align}
        \label{free_non-adjoint_estate}
        \Psi_{rs}^{(0)}(q,p)&=\frac{1}{\sqrt{2^{r+s}r!s!}}\qty(\frac{\alpha\beta H^2}{\pi^2\sigma_{qq}^{2}\sigma_{pp}^{2}})^{1/4}H_s\qty(\sqrt{\frac{\alpha H}{\sigma_{qq}^{2}}}q)H_r\qty(\sqrt{\frac{\beta H}{\sigma_{pp}^{2}}}p)e^{-\frac{\alpha H}{\sigma_{qq}^{2}}q^2-\frac{\beta H}{\sigma_{pp}^{2}}p^2},\\
        \label{free_adjoint_estate}
        \Psi_{rs}^{(0)*}(q,p)&=\frac{1}{\sqrt{2^{r+s}r!s!}}\qty(\frac{\alpha\beta H^2}{\pi^2\sigma_{qq}^{2}\sigma_{pp}^{2}})^{1/4}H_s\qty(\sqrt{\frac{\alpha H}{\sigma_{qq}^{2}}}q)H_r\qty(\sqrt{\frac{\beta H}{\sigma_{pp}^{2}}}p).
    \end{align}
\end{subequations}
For the case where $\sigma_{pp}^2=0$, the eigenstates can be written as\footnote{To take the limit, we have used the identity $\lim_{\epsilon\rightarrow0}\frac{(-1)^{-n}(\sqrt{2}\epsilon)^{n-1}}{\sqrt{\pi}}H_n\qty(\frac{x}{\sqrt{2}\epsilon})e^{-\frac{x^2}{2\epsilon^2}}=\delta^{(n)}(x)$.}
\begin{subequations}
    \label{free_eigenstates_spp=0}
    \begin{align}
        \label{free_non-adjoint_estate_spp=0}
        \lim_{\sigma_{pp}^2\rightarrow0}\Psi_{rs}^{(0)}(q,\Tilde{p})&=\frac{(-1)^{-r}}{\sqrt{2^{r+s}r!s!}}\qty(\frac{\alpha H}{\sigma_{qq}^2})^{1/4}\delta^{(r)}(\Tilde{p})H_s\qty(\sqrt{\frac{\alpha H}{\sigma_{qq}^2}q})e^{-\frac{\alpha H}{\sigma_{qq}^2}q^2},\\
        \label{free_adjoint_estate_spp=0}
        \lim_{\sigma_{pp}^2\rightarrow0}\Psi_{rs}^{(0)*}(q,\Tilde{p})&=\sqrt{\frac{2^r}{2^sr!s!}}\qty(\frac{\alpha H}{\pi^2\sigma_{qq}^2})^{1/4}\Tilde{p}^rH_s\qty(\sqrt{\frac{\alpha H}{\sigma_{qq}^2}q}),
    \end{align}
\end{subequations}
where $\Tilde{p}=\sqrt{\frac{\beta H}{\sigma_{pp}^2}}p$ and superscript $(r)$ indicates we are taking the $r$th derivative of the $\delta$-function. These are well behaved eigenstates if we use $(q,\Tilde{p})$ as our variables, with which we have the biorthogonality and completeness relations.  

\subsubsection{Stochastic 2-point functions to $\mathcal{O}(\lambda_S)$}
\label{subsubsec:stochastic_perturbation_theory}

We will now move to the more interesting case of an interacting theory. To relate the stochastic correlators to the perturbative results of QFT, we expand our solutions to the eigenproblem (\ref{FP_eigenequations}) in terms of the $(q,p)$ variables to $\mathcal{O}(\lambda_S)$
\begin{subequations}
    \label{perturbed_eigensolutions}
    \begin{align}
        \Lambda_{rs}=&\Lambda_{rs}^{(0)}+\lambda_S \Lambda_{rs}^{(1)}+\mathcal{O}(\lambda_S^2),\\
        \Psi_{rs}^{(*)}(q,p)=&\Psi_{rs}^{(0)(*)}(q,p)+\lambda_S\Psi_{rs}^{(1)(*)}(q,p)+\mathcal{O}(\lambda_S^2).
    \end{align}
\end{subequations}  
Using the eigenequations with the biorthogonality conditions for $(q,p)$, equivalent to  Eq. (\ref{FP_eigenequations}) and (\ref{biorthogonal&completeness_relations}), the $\mathcal{O}(\lambda)$ terms in the eigenvalues and eigenstates are given by
\begin{subequations}
    \label{O(lambda)_eigenvalues&eigenstates}
    \begin{align}
    \label{O(lambda)_eigenvalues}
        \Lambda_{rs}^{(1)}&=-\qty(\Psi_{rs}^{(0)*},\mathcal{L}_{FP}^{(1)}\Psi_{rs}^{(0)}),\\
    \label{O(lambda)_eigenstates}  
        \Psi_{rs}^{(1)}(q,p)&=\sum_{r's'}\Psi_{r's'}^{(0)}(q,p)\frac{\qty(\Psi_{r's'}^{(0)*},\mathcal{L}_{FP}^{(1)}\Psi_{rs}^{(0)})}{\Lambda_{r's'}^{(0)}-\Lambda_{rs}^{(0)}},\\ 
    \label{O(lambda)_adjoint_eigenstates}   
        \Psi_{rs}^{(1)*}(q,p)&=\sum_{r's'}\Psi_{r's'}^{(0)*}(q,p)\frac{\qty(\Psi_{r's'}^{(0)},\mathcal{L}_{FP}^{(1)*}\Psi_{rs}^{(0)*})}{\Lambda_{r's'}^{(0)}-\Lambda_{rs}^{(0)}}, 
    \end{align}
\end{subequations}
where for Eq. (\ref{O(lambda)_eigenstates}) and (\ref{O(lambda)_adjoint_eigenstates}), $r'\ne r$ and $s'\ne s$. One can then compute the spacelike stochastic 2-point correlator to $\mathcal{O}(\lambda_S)$ using Eq. (\ref{2pt_spacelike_stochastic_correlators_fg}) as
\begin{equation}
\label{O(lambda)_spacelike_stochastic_correlator}
    \begin{split}
        \expval{f(q_1,p_1)g(q_2,p_2)}=&\int dq_r\int dp_r\sum_{r'rs's}\Bigg[\frac{\Psi_{00}^{(0)}(q_r,p_r)}{\Psi_{00}^{(0)*}(q_r,p_r)}\Psi_{rs}^{(0)*}(q_r,p_r)\Psi_{r's'}^{(0)*}(q_r,p_r)f_{rs}^{(0)}g_{r's'}^{(0)}\\&
        \begin{split}
        +\lambda\Bigg(&\frac{\Psi_{00}^{(1)}(q_r,p_r)}{\Psi_{00}^{(0)*}(q_r,p_r)}\Psi_{rs}^{(0)*}(q_r,p_r)\Psi_{r's'}^{(0)*}(q_r,p_r)f_{rs}^{(0)}g_{r's'}^{(0)}
        \\&-\frac{\Psi_{00}^{(1)*}(q_r,p_r)\Psi_{00}^{(0)}(q_r,p_r)}{\Psi_{00}^{(0)*}(q_r,p_r)^2}\Psi_{rs}^{(0)*}(q_r,p_r)\Psi_{r's'}^{(0)*}(q_r,p_r)f_{rs}^{(0)}g_{r's'}^{(0)}
        \\&+\frac{\Psi_{00}^{(0)}(q_r,p_r)}{\Psi_{00}^{(0)*}(q_r,p_r)}\Psi_{rs}^{(1)*}(q_r,p_r)\Psi_{r's'}^{(0)*}(q_r,p_r)f_{rs}^{(0)}g_{r's'}^{(0)}
        \\&+\frac{\Psi_{00}^{(0)}(q_r,p_r)}{\Psi_{00}^{(0)*}(q_r,p_r)}\Psi_{rs}^{(0)*}(q_r,p_r)\Psi_{r's'}^{(1)*}(q_r,p_r)f_{rs}^{(0)}g_{r's'}^{(0)}
        \\&+\frac{\Psi_{00}^{(0)}(q_r,p_r)}{\Psi_{00}^{(0)*}(q_r,p_r)}\Psi_{rs}^{(0)*}(q_r,p_r)\Psi_{r's'}^{(0)*}(q_r,p_r)f_{rs}^{(1)}g_{r's'}^{(0)}
        \\&+\frac{\Psi_{00}^{(0)}(q_r,p_r)}{\Psi_{00}^{(0)*}(q_r,p_r)}\Psi_{rs}^{(0)*}(q_r,p_r)\Psi_{r's'}^{(0)*}(q_r,p_r)f_{rs}^{(0)}g_{r's'}^{(1)}\Bigg)\Bigg]
        \end{split}
        \\&\times \abs{Ha(t)(\mathbf{x}_1-\mathbf{x}_2)}^{-\qty(\frac{\Lambda_{rs}^{(0)}+\Lambda_{r's'}^{(0)}}{H}+\lambda\frac{\Lambda_{rs}^{(1)}+\Lambda_{r's'}^{(1)}}{H})}.
    \end{split}
\end{equation}
We also wish to incorporate $\mathcal{O}(\lambda_S)$ effects into our stochastic parameters so we expand them about the free parameters (\ref{free_matched_stochastic_parameters}) such that
\begin{subequations}
    \label{perturbative_expansion_stochastic_parameters}
    \begin{align}
    m_S^2&=m_R^2+\lambda m_S^{2(1)},\\
    \sigma_{qq}^2&=\frac{H^3\lb3/2-\nu_R\rb\nu_R}{4\pi^2\lb3/2+\nu_R\rb}\frac{\Gamma(2\nu_R)\Gamma\lb\frac{3}{2}-\nu_R\rb4^{\frac{3}{2}-\nu_R}}{\Gamma\lb\frac{1}{2}+\nu_R\rb}+\lambda_S\sigma_{qq}^{2(1)}+\mathcal{O}(\lambda_S^2),\\
    \sigma_{qp}^2&=\lambda\sigma_{qp}^{2(1)}+\mathcal{O}(\lambda_S^2),\\
    \sigma_{pp}^2&=\lambda\sigma_{pp}^{2(1)}+\mathcal{O}(\lambda_S^2).
    \end{align}
\end{subequations}
Note that because we are now considering an interacting theory, we have to use the renormalised mass $m_R$, given in Eq. (\ref{renormalised_mass_correction}). Additionally, for simplicity, we will consider the case where $\sigma_{pp}^{2(0)}=0$ for this section, though we will compare with the other case later. 

Using Eq. (\ref{2pt_spacelike_stochastic_correlators_fg}), we can compute the spacelike $q-q$, $q-p$, $p-q$ and $p-p$ stochastic 2-point functions. Converting these to $(\phi,\pi)$ variables using Eq. (\ref{(q,p)=A(phi,pi)}), we then write the spacelike stochastic field 2-point function to $\mathcal{O}(\lambda_S)$ as
\begin{equation}
    \label{phi-phi_spacelike_stochastic_correlator}
    \begin{split}
        \langle\phi(t&,\mathbf{0})\phi(t,\mathbf{x})\rangle\\=&\Bigg[\frac{H^2}{16\pi^2}\frac{\Gamma\qty(\frac{3}{2}-\nu_R)\Gamma\qty(2\nu_R)4^{\frac{3}{2}-\nu_R}}{\Gamma\qty(\frac{1}{2}+\nu_R)}+\lambda\Bigg(\frac{(3+2\nu_R)\sigma_{qq}^{2(1)}}{4\nu_R H(3-2\nu_R)}\\&+\frac{3(3-4\nu_R)H^4\Gamma(\nu_R)^2\Gamma\qty(\frac{3}{2}-\nu_R)^2}{32\pi^5\nu_R m^2}\Bigg)\Bigg]\abs{Ha(t)\mathbf{x}}^{-\frac{2\Lambda_{01}}{H}}
        \\&+\frac{\lambda \sigma_{pp}^{2(1)}}{H^3\nu_R(3+2\nu_R)^2}\abs{Ha(t)\mathbf{x}}^{-\frac{2\Lambda_{10}}{H}}-
        \lambda\qty(\frac{\sigma_{qp}^{2(1)}}{3H^2\nu_R}+\frac{H^4\Gamma(\nu_R)^2\Gamma\qty(\frac{5}{2}-\nu_R)^2}{8\pi^5\nu_R m^2})\\&\times\abs{Ha(t)\mathbf{x}}^{-3},
    \end{split}
\end{equation}
where the exponents are
\begin{subequations}
    \label{2-point_exponents}
    \begin{align}
        \Lambda_{01}=&\alpha_S H+\lambda_S\frac{3H\beta_R^3}{32\pi^2\nu_R\beta_R^3}\frac{\Gamma(2\nu_R)\Gamma\lb\frac{3}{2}-\nu_R\rb4^{\frac{3}{2}-\nu_R}}{\Gamma\lb\frac{1}{2}+\nu_R\rb}+\mathcal{O}(\lambda_S^2),
        \\
        \Lambda_{10}=&\beta_SH-\lambda_S\frac{3H\beta_R^3}{32\pi^2\nu_R\beta_R^3}\frac{\Gamma(2\nu_R)\Gamma\lb\frac{3}{2}-\nu_R\rb4^{\frac{3}{2}-\nu_R}}{\Gamma\lb\frac{1}{2}+\nu_R\rb}+\mathcal{O}(\lambda_S^2)
    \end{align}
\end{subequations}
and we have used the free matched stochastic parameters (\ref{free_matched_stochastic_parameters}).

\subsubsection{Stochastic 4-point functions to $\mathcal{O}(\lambda_S)$}

Using the perturbative eigenspectrum computed above, we can also compute the 4-point function to $\mathcal{O}(\lambda_S)$. We will focus on the connected piece to compare with its QFT counterpart (\ref{connected_4-pt_func_QFT}) so we can use the free stochastic parameters from Eq. (\ref{free_matched_stochastic_parameters}) and, for simplicity, we will make the choice $\sigma_{Q,pp}^2=0$.

In general, the equal-space stochastic 4-point functions are given by Eq. (\ref{4-pt_stochastic_function_general}). We will again switch from $(\phi,\pi)$ variables to $(q,p)$ using Eq. (\ref{(q,p)=A(phi,pi)}). The only non-zero 4-point functions that are relevant are 
\begin{subequations}
    \label{(q,p)_4-point_functions}
    \begin{align}
        \begin{split}
            \expval{q(t_1)q(t_2)q(t_3)q(t_4)}=&q_{00,01}q_{01,00}q_{00,01}q_{01,00}e^{-\Lambda_{01}(t_2-t_1)-\Lambda_{00}(t_3-t_2)-\Lambda_{01}(t_4-t_3)}
            \\&+q_{00,01}q_{01,02}q_{02,01}q_{01,00}e^{-\Lambda_{01}(t_2-t_1)-\Lambda_{02}(t_3-t_2)-\Lambda_{01}(t_4-t_3)}
            \\&+q_{00,03}q_{03,02}q_{02,01}q_{01,00}e^{-\Lambda_{03}(t_2-t_1)-\Lambda_{02}(t_3-t_2)-\Lambda_{01}(t_4-t_3)}
            \\&+q_{00,01}q_{01,02}q_{02,03}q_{03,00}e^{-\Lambda_{01}(t_2-t_1)-\Lambda_{02}(t_3-t_2)-\Lambda_{03}(t_4-t_3)},
        \end{split}\\
        \begin{split}
            \expval{p(t_1)q(t_2)q(t_3)q(t_4)}=&p_{00,01}q_{01,00}q_{00,01}q_{01,00}e^{-\Lambda_{01}(t_2-t_1)-\Lambda_{00}(t_3-t_2)-\Lambda_{01}(t_4-t_3)}
            \\&+p_{00,01}q_{01,02}q_{02,01}q_{01,00}e^{-\Lambda_{01}(t_2-t_1)-\Lambda_{02}(t_3-t_2)-\Lambda_{01}(t_4-t_3)}
            \\&+p_{00,03}q_{03,02}q_{02,01}q_{01,00}e^{-\Lambda_{03}(t_2-t_1)-\Lambda_{02}(t_3-t_2)-\Lambda_{01}(t_4-t_3)}
            \\&+p_{00,01}q_{01,02}q_{02,03}q_{03,00}e^{-\Lambda_{01}(t_2-t_1)-\Lambda_{02}(t_3-t_2)-\Lambda_{03}(t_4-t_3)},
        \end{split}\\
        \begin{split}
            \expval{q(t_1)p(t_2)q(t_3)q(t_4)}=&q_{00,01}p_{01,00}q_{00,01}q_{01,00}e^{-\Lambda_{01}(t_2-t_1)-\Lambda_{00}(t_3-t_2)-\Lambda_{01}(t_4-t_3)}
            \\&+q_{00,01}p_{01,02}q_{02,01}q_{01,00}e^{-\Lambda_{01}(t_2-t_1)-\Lambda_{02}(t_3-t_2)-\Lambda_{01}(t_4-t_3)}
            \\&+q_{00,03}p_{03,02}q_{02,01}q_{01,00}e^{-\Lambda_{03}(t_2-t_1)-\Lambda_{02}(t_3-t_2)-\Lambda_{01}(t_4-t_3)}
            \\&+q_{00,01}p_{01,02}q_{02,03}q_{03,00}e^{-\Lambda_{01}(t_2-t_1)-\Lambda_{02}(t_3-t_2)-\Lambda_{03}(t_4-t_3)},
        \end{split}\\
        \begin{split}
            \expval{q(t_1)q(t_2)p(t_3)q(t_4)}=&q_{00,01}q_{01,00}p_{00,01}q_{01,00}e^{-\Lambda_{01}(t_2-t_1)-\Lambda_{00}(t_3-t_2)-\Lambda_{01}(t_4-t_3)}
            \\&+q_{00,01}q_{01,02}p_{02,01}q_{01,00}e^{-\Lambda_{01}(t_2-t_1)-\Lambda_{02}(t_3-t_2)-\Lambda_{01}(t_4-t_3)}
            \\&+q_{00,03}q_{03,02}p_{02,01}q_{01,00}e^{-\Lambda_{03}(t_2-t_1)-\Lambda_{02}(t_3-t_2)-\Lambda_{01}(t_4-t_3)}
            \\&+q_{00,01}q_{01,02}p_{02,03}q_{03,00}e^{-\Lambda_{01}(t_2-t_1)-\Lambda_{02}(t_3-t_2)-\Lambda_{03}(t_4-t_3)},
        \end{split}\\
        \begin{split}
            \expval{q(t_1)q(t_2)q(t_3)p(t_4)}=&q_{00,01}q_{01,00}q_{00,01}p_{01,00}e^{-\Lambda_{01}(t_2-t_1)-\Lambda_{00}(t_3-t_2)-\Lambda_{01}(t_4-t_3)}
            \\&+q_{00,01}q_{01,02}q_{02,01}p_{01,00}e^{-\Lambda_{01}(t_2-t_1)-\Lambda_{02}(t_3-t_2)-\Lambda_{01}(t_4-t_3)}
            \\&+q_{00,03}q_{03,02}q_{02,01}p_{01,00}e^{-\Lambda_{03}(t_2-t_1)-\Lambda_{02}(t_3-t_2)-\Lambda_{01}(t_4-t_3)}
            \\&+q_{00,01}q_{01,02}q_{02,03}p_{03,00}e^{-\Lambda_{01}(t_2-t_1)-\Lambda_{02}(t_3-t_2)-\Lambda_{03}(t_4-t_3)}.
        \end{split}
    \end{align}
\end{subequations}
Using the free eigenspectrum (\ref{free_eigenvalues}) and (\ref{free_eigenstates_spp=0}), one can compute these explicitly as
\begin{subequations}
    \label{(q,p)_4-point_functions_explicit}
    \begin{align}
        \begin{split}
            \expval{q(t_1)q(t_2)q(t_3)q(t_4)}=&\lb\lb\frac{\sigma_{Q,qq}^2}{2\alpha H}\rb^2-\frac{6\lambda_S\beta}{H^2\alpha(\alpha-\beta)^2}\lb\frac{\sigma_{Q,qq}^2}{2\alpha H}\rb^3\rb e^{-\alpha H(t_4+t_2-t_3-t_1)}
            \\&+\lb2\lb\frac{\sigma_{Q,qq}^2}{2\alpha H}\rb^2-\frac{24\lambda_S\beta}{H^2\alpha(\alpha-\beta)^2}\lb\frac{\sigma_{Q,qq}^2}{2\alpha H}\rb^3\rb e^{-\alpha H(t_4+t_3-t_2-t_1)}
            \\&+\frac{3\lambda_S\beta(\sigma_{Q,qq}^2)^3}{32H^5\alpha^4\nu^2}e^{-\alpha H(t_4+t_3+t_2-3t_1)}\\&+\frac{3\lambda_S\beta(\sigma_{Q,qq}^2)^3}{32H^5\alpha^4\nu^2}e^{-\alpha H(3t_4-t_4-t_2-t_1)},
        \end{split}\\
        \begin{split}
            \expval{p(t_1)q(t_2)q(t_3)q(t_4)}=&-\frac{\lambda_S\beta^2}{H(\alpha-\beta)^2}\lb\frac{\sigma_{Q,qq}^2}{2\alpha H}\rb^3e^{-\alpha H(t_4+t_2-t_3-t_1)}
            \\&-\frac{2\lambda_S\beta^2}{H(\alpha-\beta)^2}\lb\frac{\sigma_{Q,qq}^2}{2\alpha H}\rb^3e^{-\alpha H(t_4+t_3-t_2-t_1)}
            \\&-\frac{3\lambda_S\beta^2(\sigma_{Q,qq}^2)^3}{32H^5\alpha^3(3\alpha+\beta)\nu^2}e^{-\alpha H(t_4+t_3+t_2-3t_1)},
        \end{split}\\
        \begin{split}
            \expval{q(t_1)p(t_2)q(t_3)q(t_4)}=&\frac{3\lambda_S\beta^2}{H(\alpha-\beta)^3}\lb\frac{\sigma_{Q,qq}^2}{2\alpha H}\rb^3e^{-\alpha H(t_4+t_2-t_3-t_1)}
            \\&-\frac{4\lambda_S\beta^2}{H(\alpha-\beta)^2}\lb\frac{\sigma_{Q,qq}^2}{\alpha H}\rb^3e^{-\alpha H(t_4+t_3-t_2-t_1)},
        \end{split}\\
        \begin{split}
            \expval{q(t_1)q(t_2)p(t_3)q(t_4)}=&\frac{3\lambda_S\beta^2}{H(\alpha-\beta)^3}\lb\frac{\sigma_{Q,qq}^2}{\alpha H}\rb^3e^{-\alpha H(t_4+t_2-t_3-t_1)}
            \\&+\frac{12\lambda_S\beta^2}{H(\alpha-\beta)^3}\lb\frac{\sigma_{Q,qq}^2}{2\alpha H}\rb^3 e^{-\alpha H(t_4+t_3-t_2-t_1)},
        \end{split}\\
        \begin{split}
            \expval{q(t_1)q(t_2)q(t_3)p(t_4)}=&
            -\frac{\lambda_S\beta^2}{H(\alpha-\beta)^2}\lb\frac{\sigma_{Q,qq}^2}{\alpha H}\rb^3e^{-\alpha H(t_4+t_2-t_3-t_1)}
            \\&+\frac{6\lambda_S\beta^2}{H(\alpha-\beta)^3}\lb\frac{\sigma_{Q,qq}^2}{2\alpha H}\rb^3 e^{-\alpha H(t_4+t_3-t_2-t_1)}
            \\&+\frac{3\lambda_S\beta^3(\sigma_{Q,qq}^2)^3}{16H^4\alpha^3(3\alpha-\beta)\nu^2}e^{-\alpha H(3t_4-t_3-t_2+t_1)}.
        \end{split}
    \end{align}
\end{subequations}
We can then compute the timelike stochastic field 4-point function by converting back to $(\phi,\pi)$ variables to get
\begin{equation}
    \label{field_stochastic_4pt_func}
    \begin{split}
    \expval{\phi(t_1)\phi(t_2)\phi(t_3)\phi(t_4)}=&
    \frac{\lambda_S\beta^3(2\beta^2+21\beta-72)(\sigma_{Q,qq}^2)^3}{4H^5\alpha^4(\alpha-\beta)^5}e^{-\alpha H(t_2-t_1+t_4-t_3)}
    \\&+\frac{3\lambda_S\beta^3(\beta^2+\beta-6)(\sigma_{Q,qq}^2)^3}{2H^5\alpha^4(\alpha-\beta)^5}e^{-\alpha H(t_3-t_1+t_4-t_2)}
    \\&+\frac{3\lambda_S\beta^3(4\alpha+\beta)(\sigma_{Q,qq}^{2})^3}{128H^5\alpha^4(3\alpha+\beta)\nu^4}e^{-\alpha H(t_4+t_3+t_2-3t_1)}\\&-\frac{3\lambda_S\beta^3(\sigma_{Q,qq}^{2})^3}{64H^5\alpha^4(3\alpha-\beta)\nu^3}e^{-\alpha H(3t_4-t_3-t_2+t_1)}.
    \end{split}
\end{equation}
In a similar definition to the QFT, we can define the connected stochastic 4-point function as
\begin{equation}
    \label{stochastic_connected_4-pt_func_def}
    \begin{split}
    \expval{\phi(t_1)\phi(t_2)\phi(t_3)\phi(t_4)}_C=&\expval{\phi(t_1)\phi(t_2)\phi(t_3)\phi(t_4)}
    \\&-\expval{\phi(t_1)\phi(t_2)}\expval{\phi(t_3)\phi(t_4)}
    \\&-\expval{\phi(t_1)\phi(t_3)}\expval{\phi(t_2)\phi(t_4)}
    \\&-\expval{\phi(t_1)\phi(t_4)}\expval{\phi(t_2)\phi(t_3)},
    \end{split}
\end{equation}
Thus, we can write the timelike stochastic connected field 4-point function as
\begin{equation}
    \label{stochastic_timelike_4pt_func}
    \begin{split}
    \expval{\phi(t_1)\phi(t_2)\phi(t_3)\phi(t_4)}_C=&
    \frac{\lambda_S\beta^3(2\beta^2+21\beta-81)(\sigma_{Q,qq}^2)^3}{4H^5\alpha^4(\alpha-\beta)^5}e^{-\alpha H(t_2-t_1+t_4-t_3)}
    \\&+\frac{3\lambda_S\beta^3(\beta^2+\beta-9)(\sigma_{Q,qq}^2)^3}{2H^5\alpha^4(\alpha-\beta)^5}e^{-\alpha H(t_3-t_1+t_4-t_2)}
    \\&+\frac{3\lambda_S\beta^3(4\alpha+\beta)(\sigma_{Q,qq}^{2})^3}{128H^5\alpha^4(3\alpha+\beta)\nu^4}e^{-\alpha H(t_4+t_3+t_2-3t_1)}\\&-\frac{3\lambda_S\beta^3(\sigma_{Q,qq}^{2})^3}{64H^5\alpha^4(3\alpha-\beta)\nu^3}e^{-\alpha H(3t_4-t_3-t_2+t_1)}.
    \end{split}
\end{equation}
We can also compute the equal-time stochastic connected 4-point function at $\mathcal{O}(\lambda_S)$. Using Eq. (\ref{spacelike_4-pt_correlator_f}), and recalling that $\abs{\mathbf{x}}=\abs{\mathbf{x}_j-\mathbf{x}_i}$ $\forall i\ne j$, we find that the only non-zero spacelike $(q,p)$ correlators that are relevant are
\begin{subequations}
    \label{spacelike_(q,p)_4-pt_functions}
    \begin{align}
        \begin{split}
        \expval{q(\mathbf{x}_1)q(\mathbf{x}_2)q(\mathbf{x}_3)q(\mathbf{x}_4)}=&\int dq_r\int dp_r \frac{\Psi_{0}(q_r,p_r)}{\Psi_0^*(q_r,p_r)^3}\lb\Psi_1^*(q_r,p_r)q_{10}\rb^4\abs{Ha(t)\mathbf{x}}^{-\frac{4\Lambda_1}{H}}
        \\&+\int dq_r\int dp_r \frac{\Psi_{0}(q_r,p_r)}{\Psi_0^*(q_r,p_r)^3}\lb\Psi_1^*(q_r,p_r)q_{10}\rb^3\lb\Psi_3^*(q_r,p_r)q_{30}\rb\\&\times\abs{Ha(t)\mathbf{x}}^{-\frac{3\Lambda_1+\Lambda_3}{H}},
        \end{split}\\
        \begin{split}
        \expval{q(\mathbf{x}_1)q(\mathbf{x}_2)q(\mathbf{x}_3)p(\mathbf{x}_4)}=&\expval{q(\mathbf{x}_1)q(\mathbf{x}_2)p(\mathbf{x}_3)q(\mathbf{x}_4)}
        \\=&\expval{q(\mathbf{x}_1)p(\mathbf{x}_2)q(\mathbf{x}_3)q(\mathbf{x}_4)}
        \\=&\expval{p(\mathbf{x}_1)q(\mathbf{x}_2)q(\mathbf{x}_3)q(\mathbf{x}_4)}
        \\=&\int dq_r\int dp_r \frac{\Psi_{0}(q_r,p_r)}{\Psi_0^*(q_r,p_r)^3}\lb\Psi_1^*(q_r,p_r)q_{10}\rb^3\lb\Psi_3^*(q_r,p_r)p_{10}\rb\\&\times\abs{Ha(t)\mathbf{x}}^{-\frac{4\Lambda_1}{H}}
        \\&+\int dq_r\int dp_r \frac{\Psi_{0}(q_r,p_r)}{\Psi_0^*(q_r,p_r)^3}\lb\Psi_1^*(q_r,p_r)q_{10}\rb^3\lb\Psi_3^*(q_r,p_r)p_{30}\rb\\&\times\abs{Ha(t)\mathbf{x}}^{-\frac{3\Lambda_1+\Lambda_3}{H}}.
        \end{split}
    \end{align}
\end{subequations}
Then we can use the eigenspectrum (\ref{free_eigenvalues}) and (\ref{free_eigenstates_spp=0}) to obtain
\begin{subequations}
    \begin{align}
        \label{qqqq_spacelike_correlator}
        \begin{split}
            \expval{q(\mathbf{x}_1)q(\mathbf{x}_2)q(\mathbf{x}_3)q(\mathbf{x}_4)}
            =&\lb\frac{3(\sigma_{Q,qq}^{2})^2}{4H^2\alpha^2}-\frac{3\lambda_S\beta(\sigma_{Q,qq}^2)^3}{H^5\alpha^4(\alpha-\beta)^2}\rb\abs{Ha(t)\mathbf{x}}^{-4\alpha}
            \\&+\frac{3\lambda_S\beta\lb\sigma_{Q,qq}^{2}\rb^3}{2H^5\alpha^4(\alpha-\beta)^2}\abs{Ha(t)\mathbf{x}}^{-6\alpha},
        \end{split}\\
        \begin{split}
            \expval{p(\mathbf{x}_1)q(\mathbf{x}_2)q(\mathbf{x}_3)q(\mathbf{x}_4)}
            =&\frac{9\lambda_S\beta^2\lb\sigma_{Q,qq}^2\rb^3}{8H^4\alpha^3(\alpha-\beta)^3}\abs{Ha(t)\mathbf{x}}^{-4\alpha}
            \\&+\frac{3\lambda_S\beta^2\lb\sigma_{Q,qq}^{2}\rb^3}{4H^4\alpha^3(\alpha-\beta)^2(3\alpha-\beta)}\abs{Ha(t)\mathbf{x}}^{-6\alpha}.
        \end{split}
    \end{align}
\end{subequations}
Thus, the spacelike stochastic field 4-point function is
\begin{equation}
    \label{stochastic_4pt_func}
    \begin{split}
        \expval{\phi(\mathbf{x}_1)\phi(\mathbf{x}_2)\phi(\mathbf{x}_3)\phi(\mathbf{x}_4)}=&\lb\frac{3\beta^2\lb\sigma_{Q,qq}^2\rb^2}{4H^2\alpha^2(\alpha-\beta)^2}-\frac{9\lambda_S\beta^3(2\alpha-\beta)\lb\sigma_{Q,qq}^2\rb^3}{2H^5\alpha^4(\alpha-\beta)^5}\rb\abs{Ha(t)\mathbf{x}}^{-4\alpha}
        \\&+\frac{3\lambda_S\beta^6(\sigma_{Q,qq}^{2})^3}{16H^5\alpha^4\nu^3(\beta-3\alpha)}\abs{Ha(t)\mathbf{x}}^{-6\alpha}.
    \end{split}
\end{equation}
Then we can use Eq. (\ref{stochastic_connected_4-pt_func_def}) (replacing $t_i$ with $\mathbf{x}_i$) to get the connected spacelike stochastic 4-pt function to $\mathcal{O}(\lambda_S)$ as
\begin{equation}
    \label{stochastic_spacelike_4pt_func}
    \begin{split}
        \expval{\phi(\mathbf{x}_1)\phi(\mathbf{x}_2)\phi(\mathbf{x}_3)\phi(\mathbf{x}_4)}_C=&-\frac{9\lambda_S\beta^3\lb\sigma_{Q,qq}^2\rb^3}{4H^5\alpha^4(\alpha-\beta)^4}\abs{Ha(t)\mathbf{x}}^{-6+4\nu}
        \\&+\frac{3\lambda_S\beta^6(\sigma_{Q,qq}^{2})^3}{16H^5\alpha^4\nu^3(\beta-3\alpha)}\abs{Ha(t)\mathbf{x}}^{-9+6\nu}.
    \end{split}
\end{equation}
In the light field limit $m\ll H$, the spacelike 4-point function is given by
\begin{equation}
    \label{spacelike_4pt_func_light_fields}
    \begin{split}
    \expval{\phi(\mathbf{x}_1)\phi(\mathbf{x}_2)\phi(\mathbf{x}_3)\phi(\mathbf{x}_4)}_C\eval_{m^2\ll H^2}=&\frac{81\lambda_S H^{12}}{128\pi^6m^6}\abs{Ha(t)\mathbf{x}}^{-\frac{2m^2}{H^2}}
    \\&-\frac{243\lambda_SH^{12}}{256\pi^6m^8}\abs{Ha(t)\mathbf{x}}^{-\frac{4m^2}{3H^2}},
    \end{split}
\end{equation}
which is the same as the near-massless QFT spacelike 4-point function (\ref{spacelike_quantum_4-pt_func_light_fields}) for $\lambda_S=\lambda$.

\subsubsection{Stochastic parameters to $\mathcal{O}(\lambda)$}
\label{subsec:perturbative_stochastic_parameters}

We can now compute the $\mathcal{O}(\lambda)$ stochastic parameters by comparing the $\mathcal{O}(\lambda_S)$ stochastic 2-point and 4-point functions with their QFT counterparts. The procedure for matching our 2-point functions remains unchanged from the free case; we again have our three conditions (plus a choice) for the stochastic results to match perturbative QFT: (i)  the leading exponents match, (ii) the leading prefactors match, (iii) the analytic continuation between spacelike and timelike correlators is preserved and (iv) either the subleading terms match or it vanishes in the stochastic correlators. We will consider both cases presented by (iv), though as stated these are not unique choices. For the following, I will consider the latter choice; the former is dealt with in Appendix \ref{app:stochastic_parameters_NLO}. We will see that the choice doesn't affect the physical results, as indeed it shouldn't.

However, we now have a fifth stochastic parameter to contend with - $\lambda_S$ - which can be matched to its QFT counterpart by considering the connected 4-point function. By equating the spacelike stochastic 4-point function (\ref{stochastic_spacelike_4pt_func}), using the free stochastic parameters (\ref{free_matched_stochastic_parameters}), to the spacelike quantum 4-point function (\ref{spacelike_quantum_4-pt_func}), we find that the term $\mathcal{O}(\abs{Ha(t)\mathbf{x}}^{-9+6\nu}$ are equal for any value of $m$ if
\begin{equation}
    \label{matched_stochastic_lambda}
    \lambda_S=\lambda+\mathcal{O}(\lambda^2).
\end{equation}
Thus, at this order, the $\lambda$ parameters in both quantum and stochastic theories are the same and we will drop the subscript $S$ henceforth. We note that the second-order stochastic theory also gives us an expression of $\mathcal{O}\lb\abs{Ha(t)\mathbf{x}}^{-6+4\nu}\rb$, which we know also appears in the QFT counterpart (\ref{spacelike_quantum_4-pt_func}), denoted by the `...+'. Thus, the stochastic theory gives us a way of computing this term explicitly, which is difficult to do in perturbative QFT.

We can now turn our attention to the 2-point functions, and the other 4 $\mathcal{O}(\lambda)$ stochastic parameters. In order for the stochastic spacelike 2-point function (\ref{phi-phi_spacelike_stochastic_correlator}) to reproduce the perturbative QFT 2-point function (\ref{O(lambda)_2-pt_function_spacelike_long-distance}), the parameters must have the values
\begin{subequations}
    \label{one-loop_qp_stochastic_parameters}
    \begin{align}
        \label{one-loop_stochastic_mass}
        \begin{split}
         m_S^2=&m_R^2+\frac{3\lambda H^2}{16\pi^2}\Bigg[-\frac{4\Gamma\qty(3/2-\nu_R)\Gamma\qty(\nu_R)}{\sqrt{\pi}}\\&+\qty(2-\frac{m_R^2}{H^2})\qty(1-\psi^{(0)}\qty(3/2-\nu_R)-\psi^{(0)}\qty(3/2+\nu_R)+\ln\qty(\frac{M^2}{a(t)^2H^2}))\Bigg]\\&+\mathcal{O}(\lambda^2),
         \end{split}\\
        \label{one-loop_sigma_qq}
        \begin{split}        \sigma_{qq}^{2}=&\frac{2H^3\Gamma(1+\nu_R)\Gamma\qty(\frac{5}{2}-\nu_R)}{\pi^{5/2}\qty(3+2\nu_R)}+\frac{\lambda H^3\Gamma\qty(3/2-\nu_R)\Gamma\qty(\nu_R)}{16\pi^{7/2}\qty(3+2\nu_R)^2}\\&\times\Bigg[3(-3+2\nu_R)\Gamma\qty(3/2-\nu_R)\Gamma\qty(2\nu_R)+\frac{3\times4^{-3+\nu_R}}{\nu_R}\qty(-1+4\nu_R^2)\\&\times\Gamma\qty(1/2+\nu_R)\Bigg(4\frac{m_R^2}{H^2}-12\nu_R-4\frac{m^2_R}{H^2}\nu_R\ln4-4\frac{m_R^2}{H^2}\nu_R\psi^{(0)}\qty(3/2-\nu_R)\\&+8\frac{m_R^2}{H^2}\nu_R\psi^{(0)}(2\nu_R)-4\frac{m_R^2}{H^2}\nu_R\psi^{(0)}\qty(1/2+\nu_R)\Bigg)\\&\qty(-1-\ln\qty(\frac{M^2}{a(t)^2H^2})+\psi^{(0)}\qty(3/2-\nu_R)+\psi^{(0)}\qty(3/2+\nu_R))\Bigg]+\mathcal{O}(\lambda^2),
        \end{split}\\
        \label{one-loop_sigma_qp}
        \begin{split}
        \sigma_{qp}^{2}=&\frac{3\lambda H^4(-3+2\nu_R)\Gamma\qty(3/2-\nu_R)^2\Gamma\qty(\nu_R)^2}{32\nu_R\pi^5}+\mathcal{O}(\lambda^2),
        \end{split}\\
        \label{one-loop_sigma_pp}
        \begin{split}
        \sigma_{pp}^2=&\mathcal{O}(\lambda^2).
        \end{split}
    \end{align}
\end{subequations}
Note that the mass parameter is now dependent on the renormalisation scale $M$. Using these parameters in our second-order stochastic equations (\ref{2d_stochastic_eq}) gives us a second-order stochastic theory of quartic self-interacting scalar QFT in de Sitter. 

These expressions still have an IR problem, but it is milder than that of perturbative QFT. Expanding the $\mathcal{O}(\lambda)$ terms to leading order in $m^2/H^2$, we have
\begin{subequations}
    \label{O(lambdaH2m2)_stochastic_parameters}
    \begin{align}
        m_S^{2}=&m_R^2+\frac{3\lambda H^2}{8\pi^2}\qty(2\gamma_E-\ln4+\ln\qty(\frac{M^2}{a(t)^2H^2}))+\mathcal{O}\qty(\lambda m_R^2),\\
        \sigma_{qq}^2=&\frac{2H^3\Gamma(1+\nu_R)\Gamma\qty(\frac{5}{2}-\nu_R)}{\pi^{5/2}\qty(3+2\nu_R)}+\frac{\lambda H^5\qty(-8+3\ln4)}{32\pi^4 m_R^2}+\mathcal{O}\qty(\lambda H^3),\\
        \sigma_{qp}^2=&-\frac{3\lambda H^6}{32\pi^4m_R^2}+\mathcal{O}\qty(\lambda H^4),\\
        \sigma_{pp}^2=&0+\mathcal{O}\qty(\lambda^2).
    \end{align}
\end{subequations}
We see that the sum will converge when $\lambda\ll m^2/H^2$; however, since we have corrected at this order, the error associated with the stochastic parameters is actually $\mathcal{O}\lb\frac{\lambda^2H^4}{m^4}\rb$. Thus, the second-order stochastic theory is limited to $\lambda^2\ll m^4/H^4$. This is a limitation of the matching procedure required to obtain the stochastic parameters, since we rely on the results of perturbative QFT. Crucially, the IR problem is less serious in our stochastic approach compared with perturbative QFT: $\mathcal{O}(\lambda^2 H^4/m^4)$ as opposed to $\mathcal{O}(\lambda H^4/m^4)$. 

Converting our stochastic noise back to using the $(\phi,\pi)$ variables, our stochastic parameters are given by
\begin{subequations}
\label{stochastic_parameters_O(lambdaH2m2)}
\begin{align}
    \label{stochastic_mass_O(lambdaH2m2)}
    m_S^{2}=&m_R^2+\frac{3\lambda H^2}{8\pi^2}\qty(2\gamma_E-\ln4+\ln\qty(\frac{M^2}{a(t)^2H^2}))+\mathcal{O}\qty(\lambda m_R^2)\\
    \label{stochastic_lambda}
    \lambda_S=&\lambda+\mathcal{O}(\lambda^2)\\
    \label{O(lambdaH2m2)_phi-pi_noise_matrix}
    \begin{split}
    \sigma^{2}=&\frac{H^3\Gamma\qty(\nu_R)\Gamma\qty(\frac{5}{2}-\nu_R)}{2\pi^{5/2}}\begin{pmatrix}
        1&-\frac{2m_R^2}{H(3+2\nu_R)}\\-\frac{2m_R^2}{H(3+2\nu_R)}&\frac{4m_R^4}{(3+2\nu_R)^2H^2}
    \end{pmatrix}
    \\&+\lambda
    \begin{pmatrix}
    \frac{3H^5\qty(-2+\ln4)}{32\pi^4m_R^2}+\mathcal{O}\qty(H^3) & -\frac{3H^6}{32\pi^4m_R^2}+\mathcal{O}\qty(H^4) \\ -\frac{3H^6}{32\pi^4m_R^2}+\mathcal{O}\qty(H^4) & \mathcal{O}\qty(H^5)
    \end{pmatrix}.
    \end{split}
\end{align}
\end{subequations}
Thus, using the stochastic parameters (\ref{stochastic_parameters_O(lambdaH2m2)}) elevates the stochastic theory (\ref{2d_stochastic_eq}) to an IR effective theory of quartic self-interacting scalar QFT in de Sitter.

\section{Comparison with other approximations}
\label{sec:comparison}

\subsection{Regimes of validity}
\label{subsec:regimes_of_validity}

One of the strengths of the stochastic approach is that one can employ numerical techniques to compute correlation functions. For the second-order theory, one can use a matrix diagonalisation scheme to compute the eigenspectrum non-perturbatively. This method is outlined in Ref. \cite{Cable:2021}. Thus, the only limitation is the perturbative computation of the stochastic parameters. Conversely, the perturbative method outlined in Sec. \ref{sec:scalar_QFT} for QFT doesn't have an associated non-perturbative scheme; all results are purely perturbative. Thus, the second-order stochastic theory can be used to go beyond perturbative QFT. Additionally, the second-order theory extends the regime of validity of the overdamped stochastic theory pioneered by Starobinsky and Yokoyama \cite{starobinsky:1986,Starobinsky-Yokoyama:1994}. For more details on this method, see Ref. \cite{Cable:2021,Markkanen:2019}.

\begin{figure}[ht]
    \centering
    \includegraphics[width=150mm]{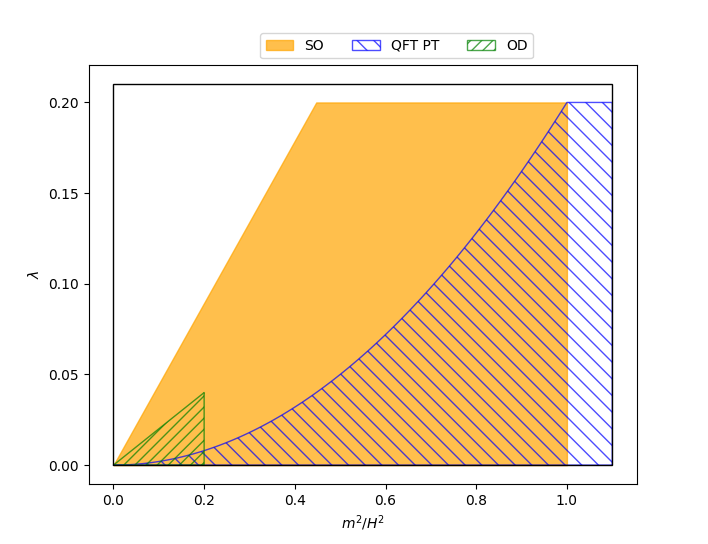}
    \caption{This shows the regimes in which we expect our approximations to work. Perturbative QFT, OD stochastic and second-order (SO) stochastic are expected to work in the blue left hashed, green right hashed and orange regions respectively. Note that there is some overlap. The pure white space is where none of these approximations work.}
    \label{fig:model_limitations_comparison}
\end{figure}

For the massive, quartic self-interacting scalar field theory considered here, the regime of these three approximations is
\begin{subequations}
    \label{model_limitations}
    \begin{align}
        \text{Perturbative QFT: }\qquad\lambda\ll &\frac{m^4}{H^4}, \qquad \lambda\ll1,\\
        \text{OD stochastic: }\qquad\lambda\ll&\frac{m^2}{H^2}, \qquad m\ll H, \\
        \text{Second-order stochastic: }\qquad\lambda^2\ll &\frac{m^4}{H^4}, \qquad m\lesssim H.
    \end{align}
\end{subequations}
We make a graphical comparison of these regimes in Fig. \ref{fig:model_limitations_comparison}. For the purposes of  making the boundaries obvious, we choose ``$\ll1$'' to mean ``$<0.2$'', though in reality we wouldn't expect these boundaries to be so clear cut. 

The blue left hashed region represents the parameter space described by perturbative QFT. We can see that for light fields $m\lesssim H$, this region is entirely covered by the second-order stochastic theory. This is unsurprising given that the stochastic correlators were found directly from the 2-point functions of perturbative QFT. Beyond the light field limit, perturbative QFT continues to extend (though it is still limited to $\lambda\ll1$ - it is after all a perturbative theory!). This extension is not covered by either stochastic approaches as they both require light fields. 

The overdamped stochastic approach - the green, right hashed region - is resigned to near-massless $m\ll H$ fields, but does go beyond perturbative QFT due to the non-perturbative methods available to it. Further, it is far simpler to compute stochastic correlation functions than their QFT counterparts, hence its popularity within its regime of validity. 

The OD stochastic approach is encompassed by the second-order stochastic effective theory, as represented by the orange region in Fig. \ref{fig:model_limitations_comparison}.  However, the second-order stochastic effective theory goes further, also encompassing perturbative QFT entirely in the light field limit. We can also see that there is a large chunk of the parameter space, even for near-massless fields, that is only covered by the second-order stochastic theory. The introduction of $\mathcal{O}(\lambda)$ corrections to the stochastic parameters means it goes beyond the OD approach, even in the limit $m\ll H$, while the non-perturbative methods available mean that it can extend beyond perturbative QFT\footnote{Note that, due to the matching procedure, it is still limited to the region $\lambda\ll 1$ as the stochastic parameters are found perturbatively.}. This suggests that the second-order stochastic effective theory can be used to probe hitherto untapped regions of the parameter space.

\subsection{Comparing approximations with $\mathcal{O}(\lambda)$ stochastic parameters}
\label{sec:comparing_approx}

In Ref. \cite{Cable:2022}, we performed a careful analysis of how the three approximations compare with each other. Now, we wish to update that analysis to incorporate the updated stochastic parameters. There are two important changes that we have made here that will effect the results. The first is that we have $\mathcal{O}(\lambda)$ corrections to the stochastic parameters, which were not included in the previous paper \cite{Cable:2022}. The second is that now we have done the renormalisation more carefully and so we have a renormalisation-scale dependent mass parameter $m_R(M)$, which must be chosen when doing numerical computations.

We will revisit the two examples discussed in Sec. B1 and B2 of Ref. \cite{Cable:2022} by computing the exponent for the leading term in the long-distance behaviour of the scalar 2-point functions. For QFT, this corresponds to the quantity given in Eq. (\ref{QFT_exponent}) to $\mathcal{O}\lb\frac{\lambda H^4}{m^4}\rb$ while, 
for the two stochastic approximations - OD and second-order - the quantities in question are the first-excited eigenvalues, $\Lambda_1^{(OD)}$ and $\Lambda_1^{(SO)}$, of their respective spectral expansions. They are computed numerically. Note that the result used in Ref. \cite{Cable:2022} for the OD stochastic theory is the same as the one here, other than the fact we now use $m_R(M)$ instead of $m$.

We plot $\Lambda_1$ for all three approximations as a function of the coupling $\lambda$ for fixed $m_R^2/H^2$ and scale $M=a(t)H$. The first example will be for $m_R^2/H^2=0.1$ (Fig. \ref{fig:m20.1_match_v_OD_v_QFT}), where we are in a regime where the OD stochastic approach is valid, while the second will be for $m_R^2/H^2=0.3$ (Fig. \ref{fig:m20.3_match_v_OD_v_QFT}), where we expect it to fail. In both examples, we expect perturbative QFT to hold for small $\lambda$ and fail as $\lambda$ increases, since it is in this regime that $\lambda\rightarrow m^4/H^4$. These two approximations are given by the yellow dotted (OD stochastic) and blue dashed (perturbative QFT). 

Using similar reasoning, the second-order stochastic theory should be expected to hold for small $\lambda$ and begin to fail as we increase $\lambda$ in both plots. This is a less severe failure as the breakdown is now for $\lambda^2\rightarrow m_R^4/H^4$. We will also consider how the choice of the $\sigma_{pp}^2$ noise amplitude affects our results. The red and green lines represent the choices $\sigma_{pp}^2=0$ and $\sigma_{pp}^2=\sigma_{pp}^{2(NLO)}$ respectively. For both cases, we also consider the free (dot-dashed) and $\mathcal{O}(\lambda)$ (solid) stochastic parameters, given in Eq. (\ref{free_matched_stochastic_parameters}) and (\ref{one-loop_qp_stochastic_parameters})/(\ref{O(lambdaH2m2_stochastic_parameters_NLO_match}) respectively, so we can ascertain the effect that interacting stochastic parameters have on the results. Indeed, we will see that interactions are crucial so that physical results such as $\Lambda_1$ are not dependent on our choice of $\sigma_{pp}^2$.

\subsubsection{Example 1: $m^2/H^2=0.1$}

\begin{figure}[ht]
    \centering
    \includegraphics[width=150mm]{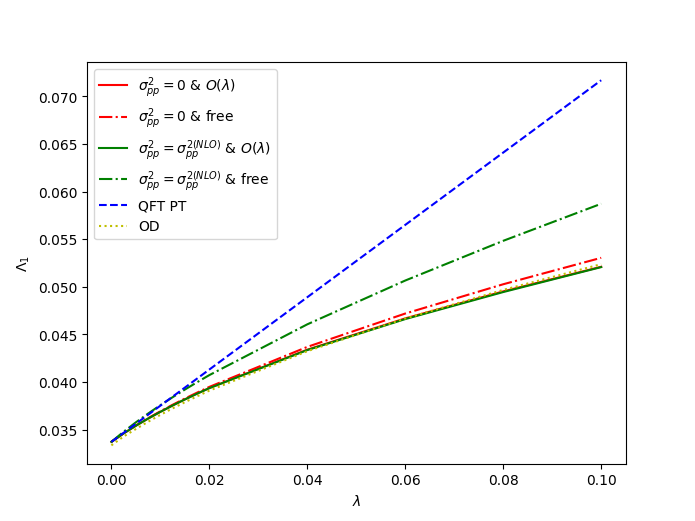}
    \caption{A plot of the first excited eigenvalue $\Lambda_1$ as a function of $\lambda$ for $m_R^2/H^2=0.1$ with the choice $M=a(t)H$, using perturbative QFT (blue, dashed), OD stochastic (yellow, dotted) and second-order stochastic approaches. Dot-dashed and solid lines indicate the second-order stochastic parameters are free (\ref{free_matched_stochastic_parameters}) and interacting (\ref{stochastic_parameters_O(lambdaH2m2)}) respectively, with the noise choice $\sigma_{pp}^2=\sigma_{pp}^{2(NLO)}$ (green)  and $\sigma_{pp}^2=0$ (red).}
    \label{fig:m20.1_match_v_OD_v_QFT}
\end{figure}

The first example is for $m_R^2/H^2=0.1$. This is chosen because the mass is sufficiently small such that the OD stochastic approach will be valid beyond perturbative QFT. Consider Fig. \ref{fig:m20.1_match_v_OD_v_QFT}. This plot is identical to that Fig. 3 in Ref. \cite{Cable:2022} other than the fact we have now also added two curves where we have introduced $\mathcal{O}(\lambda)$ effects in our stochastic parameters (solid red and green). Thus, the analysis given there about the three approximations still holds; perturbative QFT quickly breaks down as $\lambda$ increases, while the second-order and OD stochastic theories continue to agree. However, the new feature is that the new solid lines agree for all values of $\lambda$ plotted here, despite the fact that our choice of $\sigma_{pp}^2$ is different: red and green indicate $\sigma_{pp}^2=0$ and $\sigma_{pp}^2=\sigma_{pp}^{2(NLO)}$ respectively. Thus, we find that this choice doesn't affect physical results once you include the $\mathcal{O}(\lambda)$ effects, as indeed they shouldn't. However, for free stochastic parameters, this choice does affect the result; therefore, it is important to incorporate these new $\mathcal{O}(\lambda)$ effects for the theory to be reliable. 

\begin{figure}[ht]
    \centering
    \includegraphics[width=150mm]{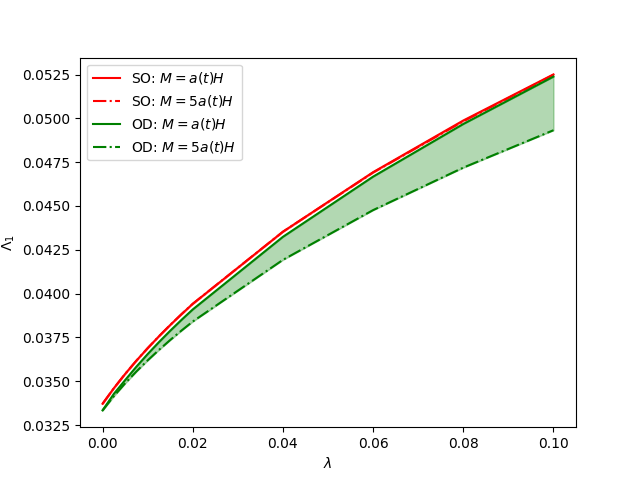}
    \caption{The first excited eigenvalue $\Lambda_1$ as a function of $\lambda$ for $m_R(a(t)H)^2/H^2=0.1$. The red and green lines show results from second-order and OD stochastic theories respectively. The solid and dot-dashed lines are for renormalisation scale choices $M=a(t)H$ and $M=5a(t)H$ respectively. The green shaded region indicates the size of the error that choosing the scale has on the OD stochastic approach. The equivalent red region is negligible because the second-order theory accounts for it via renormalisation.}
    \label{fig:m20.1_match_renorm}
\end{figure}

It is worth noting that the excellent agreement between the second-order and OD stochastic results is due to the choice of renormalisation scale $M=a(t)H$. One can see from Eq. (\ref{stochastic_mass_O(lambdaH2m2)}) that the stochastic mass $m_S$ depends on the renomalisation scale as $\sim \ln\lb\frac{M^2}{a(t)^2H^2}\rb$, which vanishes for the choice $M=a(t)H$. Thus, it is not surprising that the second-order and OD stochastic approaches agree. However, if one were to choose the renormalisation scale differently, the agreement would not be so good. For example, if one chooses $M=5 a(t)H$, the renormalised mass parameter (\ref{renormalised_mass_dim_reg_deSitter}) will no longer equal 0.1; it will have some shift of $\mathcal{O}(\lambda)$. The second-order stochastic theory accounts for this shift via the $M$-dependence in the stochastic mass $m_S$ parameter (\ref{stochastic_mass_O(lambdaH2m2)}), whereas the OD theory does not because it doesn't incorporate any UV renormalisation. Thus, the two will be different for such a choice. Fig. \ref{fig:m20.1_match_renorm} shows the effect of a different choice up to $M=5a(t)H$. One can see that there is very little change to $\Lambda_1$ for the second-order theory and that the change is much larger for the OD case. Thus, even in the regime where $m_R\ll H$, where the OD stochastic theory is deemed to be valid, it still has some error associated with UV renormalisation.

\subsubsection{Example 2: $m^2/H^2=0.3$}

\begin{figure}[ht]
    \centering
    \includegraphics[width=150mm]{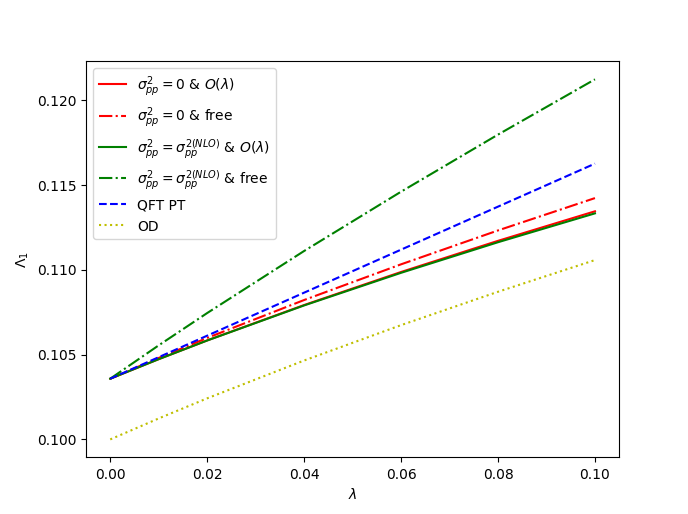}
    \caption{A plot of the first excited eigenvalue $\Lambda_1$ as a function of $\lambda$ for $m^2/H^2=0.3$ using perturbative QFT (blue, dashed), OD stochastic (yellow, dotted) and second-order stochastic approaches. Dot-dashed and solid lines indicate the second-order stochastic parameters are free and interacting respectively, with the noise choice $\sigma_{pp}^2=\sigma_{pp}^{2(NLO)}$ (green)  and $\sigma_{pp}^2=0$ (red).}
    \label{fig:m20.3_match_v_OD_v_QFT} 
\end{figure}

For completeness, we also include the example where $m_R^2/H^2=0.3$ for $M=a(t)H$, where $\Lambda_1$ as a function of $\lambda$ is given in Fig. \ref{fig:m20.3_match_v_OD_v_QFT}. This plot is identical to that of Fig. 4 in Ref. \cite{Cable:2022} except we have now included $\mathcal{O}(\lambda)$ effects to the stochastic parameters in the two new solid lines (red of $\sigma_{pp}^2=0$ and green for $\sigma_{pp}^2=\sigma_{pp}^{2(NLO)}$). Again, the analysis remains the same: the perturbative QFT diverges from the second-order as $\lambda$ increases while the OD stochastic theory never works well because the fields are too heavy. Further, in a similar way to the previous example, we see that the introduction of $\mathcal{O}(\lambda)$ effects causes the choice of $\sigma_{pp}^2$ to be irrelevant. This consolidates the point that this choice doesn't affect physical results but only when we have included the relevant $\mathcal{O}(\lambda)$ effects.

\section{Concluding remarks}
\label{sec:conclusion}

We have shown that the second-order stochastic theory is a valid effective theory of the long-distance behaviour of scalar fields in de Sitter, within the regime of validity $m\lesssim H$ and $\lambda^2\ll m^4/H^4$. This extends the work started in Ref. \cite{Cable:2021,Cable:2022} to incorporate the full $\mathcal{O}(\lambda)$ correction to the stochastic parameters, which includes a complete discussion of the UV renormalisation at this order. Notably, we have found a stochastic theory that incorporates a dependence on the renormalisation scale $M$ in such a way that physical results are $M$-independent. This is not true for the widely used overdamped stochastic approach. 

While this is the final installment in a trilogy of papers, there is plenty of study left in the second-order stochastic theory. Currently, this theory has only been tested on perturbative QFT to one-loop order. One could extend this theory to incorporate more loops, but the renormalisation procedure becomes increasingly complex. We would expect that the matching procedure used to determine the stochastic parameters would still be valid here and that the computational challenge comes from perturbative QFT. Additionally, we note that if one were to move away from equilibrium, other effects would arise for higher loops, such as secular growth \cite{Akhmedov:2013,Akhmedov:2017,Akhemdov;2019}. We have chosen to study equilibrium solutions for this paper; however, it would be interesting to consider solutions away from equilibrium to test the robustness of the second-order stochastic theory.

One could more rigorously test the effective theory by comparing it to other QFT approximations such as the $1/N$ approximation \cite{Nacir:2016,Beneke:2013} or Monte-Carlo simulations \cite{montvay:1994}. However, the main outstanding question is whether one can derive the stochastic parameters from an underlying microscopic picture, as opposed to using the matching procedure discussed here, which relies on an alternative method - in this case, perturbative QFT - being available. It is not clear to us how one should proceed in this direction.

In spite of these formal questions, the second-order effective theory already has uses in inflationary cosmology. Its numerical tools mean that it can generate novel results in this field; directions could include precision calculations of curvature and isocurvature perturbations or extensions to primordial black hole abundance computations. It is clear that this method has the potential to be an important tool in the arsenal of inflationary cosmologists. 

\acknowledgements A.C. was supported by a UK Science and Technology Facility Council studentship. A.R. was supported by STFC grants ST/T000791/1  and ST/X000575/1 and IPPP Associateship. We would like to thank Diana López Nacir, Eliel Camargo-Molina, Greg Kaplanek, Mariana Carrillo González and Sebastian Cespedes for useful discussions.

\appendix

\section{Dimensional regularisation in de Sitter spacetime}
\label{app:dim_reg_de_Sitter}

In this appendix, we compute the scalar field variance in de Sitter spacetime with Hubble rate $H$, using dimensional regularisation. This involves taking the number of spacetime dimensions to be $d=4-\epsilon$, and then taking the limit $d\rightarrow 4$.

Before we begin, it is useful to note that the calculation is straightforward with point-splitting regularisation~\cite{bunch-davies:1978}. Taking the coincident point limit of the Feynman propagator (\ref{free_feynman_propagator}) from the spacelike direction such that $x=(t,\mathbf{x})$ and $x'=(t,\mathbf{0})$, the field variance becomes
\begin{equation}
    \label{field_variance_small-x_deSitter}
    \begin{split}
    \expval{\hat{\phi}^2}_{PS}:=&i\Delta^F(\mathbf{x},\mathbf{0})\eval_{\abs{Ha(t)\mathbf{x}}\rightarrow0}\\=&-\frac{1}{4\pi^2a(t)^2\abs{\mathbf{x}}^2}
+\frac{2H^2-m_B^2}{16\pi^2}\Bigg(2\ln\frac{2}{Ha(t)\abs{\mathbf{x}}}+1-2\gamma_E\\&+\psi^{(0)}\qty(\frac{3}{2}-\nu_B)+\psi^{(0)}\qty(\frac{3}{2}+\nu_B)\Bigg),
    \end{split}
\end{equation}
where $\nu_B=\sqrt{\frac{9}{4}-\frac{m^2_B}{H^2}}$, $\psi^{(0)}(z)$ is the polygamma function and $\gamma_E$ is the Euler-Mascheroni constant. The subscript `PS' indicates it is computed using point-splitting. 

To carry out the same calculation in dimensional regularisation, we consider the $\mathbf{k}$-space integral~(\ref{2-pt_func_k-space}), which gives
\begin{equation}
    \label{k-space_field_variance}
    \begin{split}
    \expval{\hat{\phi}^2}=&\frac{\pi}{4Ha(t)^3}\int \dbar^3\mathbf{k}\abs{\mathcal{H}_{\nu}^{(1)}\lb\frac{k}{a(t)H}\rb}^2,
    \end{split}
\end{equation}
where $\dbar^3\mathbf{k}=\frac{d^3\mathbf{k}}{(2\pi)^3}$. In dimensional regularisation, the number of space dimension becomes $D=3-\epsilon$, and therefore we have
\begin{equation}
    \label{field_variance_dim_reg}
    \expval{\hat{\phi}^2}_{DR}=\frac{\pi\mu^\epsilon}{4Ha(t)^3}\int \dbar^D\mathbf{k}\abs{\mathcal{H}_{\nu}^{(1)}\lb\frac{k}{a(t)H}\rb}^2,
\end{equation}
where
$\mu$ is introduced as the regularisation scale and the subscript `DR' indicates that we have defined this integral via dimensional regularisation. This integral cannot be computed analytically, and therefore we split it in two pieces,
\begin{equation}
    \label{scalar_field_variance_k_deSitter}
    \begin{split}
    \expval{\hat{\phi}^2}_{DR}=&\mu^{\epsilon}\int\dbar^Dk\qty(\frac{1}{2a(t)^2\sqrt{k^2+\delta^2}}+\frac{2H^2-m_B^2+\frac{\delta^2}{a(t)^2}}{4\qty(k^2+\delta^2)^{3/2}})
    \\&+\int\dbar^3k\qty(\frac{\pi}{4Ha(t)^3}\abs{\mathcal{H}_{\nu}^{(1)}\qty(\frac{k}{a(t)H})}^2-\frac{1}{2a(t)^2\sqrt{k^2+\delta^2}}-\frac{2H^2-m_B^2+\frac{\delta^2}{a(t)^2}}{4\qty(k^2+\delta^2)^{3/2}}),
    \end{split}
\end{equation}
where $\delta$ is an arbitrary energy scale\footnote{Note that we cannot choose $\delta=0$ because dimensional regularisation would then give a zero result for the first line.} The sum of the two contributions will of course give a $\delta$-independent result. The first line is ultraviolet divergent but can be easily computed in dimensional regularisation, whereas the second line is finite and can therefore be computed in three dimensions. 

Computing the divergent integral in the first line of Eq. (\ref{scalar_field_variance_k_deSitter}) and taking the limit $\epsilon\rightarrow0$, we obtain
\begin{equation}
    \label{field_variance_dim_reg_deSitter}
    \begin{split}
    \expval{\hat{\phi}^2}_{DR}=&\frac{2H^2-m_B^2}{16\pi^2}\lb\frac{2}{\epsilon}-\gamma_E+\ln\lb\frac{4\pi\mu^2}{\delta^2}\rb\rb-\frac{\delta^2}{16\pi^2a(t)^2}\\&+\int\dbar^3k\Bigg(\frac{\pi}{4Ha(t)^3}\abs{\mathcal{H}_{\nu}^{(1)}\qty(\frac{k}{a(t)H})}^2-\frac{1}{2a(t)^2\sqrt{k^2+\delta^2}}-\frac{2H^2-m_B^2+\frac{\delta^2}{a(t)^2}}{4\qty(k^2+\delta^2)^{3/2}}\Bigg). 
    \end{split}
\end{equation}
The remaining integral in Eq.~(\ref{field_variance_dim_reg_deSitter}) is finite, and it can be computed using the point-splitting result in Eq.~(\ref{field_variance_small-x_deSitter}). We first note that the point-splitting regularised variance can be written as
\begin{equation}
\expval{\hat{\phi}^2}_{PS}= \frac{\pi}{4Ha(t)^3}\lim_{|\mathbf{x}|\rightarrow 0}
  \int \dbar^3\mathbf{k} e^{i\mathbf{k}\cdot\mathbf{x}}  \abs{\mathcal{H}_{\nu}^{(1)}\lb\frac{k}{a(t)H}\rb}^2.
\end{equation}
Next, we repeat the split to divergent and finite integrals in Eq.~(\ref{scalar_field_variance_k_deSitter}),
\begin{equation}
	\label{equ:phi2PS}
    \begin{split}
    \expval{\hat{\phi}^2}_{PS}=&\lim_{|\mathbf{x}|\rightarrow 0}\int\dbar^3k\ e^{i\mathbf{k}\cdot\mathbf{x}}\qty(\frac{1}{2a(t)^2\sqrt{k^2+\delta^2}}+\frac{2H^2-m_B^2+\frac{\delta^2}{a(t)^2}}{4\qty(k^2+\delta^2)^{3/2}})
    \\&+\int\dbar^3k\qty(\frac{\pi}{4Ha(t)^3}\abs{\mathcal{H}_{\nu}^{(1)}\qty(\frac{k}{a(t)H})}^2-\frac{1}{2a(t)^2\sqrt{k^2+\delta^2}}-\frac{2H^2-m_B^2+\frac{\delta^2}{a(t)^2}}{4\qty(k^2+\delta^2)^{3/2}})
\\
=&
-\frac{1}{4\pi^2a(t)^2\abs{\mathbf{x}}^2}+\frac{m_B^2-2H^2}{16\pi^2}\Bigg(
2\ln\frac{\delta\abs{\mathbf{x}}}{2}+2\gamma_E\Bigg)-\frac{\delta^2}{16\pi^2 a^2},
    \\&+\int\dbar^3k\qty(\frac{\pi}{4Ha(t)^3}\abs{\mathcal{H}_{\nu}^{(1)}\qty(\frac{k}{a(t)H})}^2-\frac{1}{2a(t)^2\sqrt{k^2+\delta^2}}-\frac{2H^2-m_B^2+\frac{\delta^2}{a(t)^2}}{4\qty(k^2+\delta^2)^{3/2}}),
    \end{split}
\end{equation}
Comparing Eqs.~(\ref{field_variance_dim_reg_deSitter}) and (\ref{equ:phi2PS}), we can see that the difference between the point-splitting and dimensional regularisation results is simply
\begin{equation}
\expval{\hat{\phi}^2}_{DR}-\expval{\hat{\phi}^2}_{PS}
=
\frac{2H^2-m_B^2}{16\pi^2}
\left(\frac{2}{\epsilon}+\ln\pi\mu^2 |\mathbf{x}|^2
+\gamma_E\right)+\frac{1}{4\pi^2a(t)^2\abs{\mathbf{x}}^2}.
\end{equation}
Using the point-splitting result from Eq.~(\ref{field_variance_small-x_deSitter}), we can therefore write the field variance in dimensional regularisation as
\begin{equation}
    \label{field_variance_dim_reg_deSitter_ana_app}
    \expval{\hat{\phi}^2}_{DR}=\frac{2H^2-m_B^2}{16\pi^2}\qty[\frac{2}{\epsilon}+\ln\frac{4\pi\mu^2}{a(t)^2H^2}-\gamma_E+1-\psi^{(0)}\qty(\frac{3}{2}-\nu_B)-\psi^{(0)}\qty(\frac{3}{2}+\nu_B)].
\end{equation}

\section{The IR behaviour of the connected four-point function}
\label{app:IR_lim_4-pt_func}

In this appendix, we will consider more carefully the IR limit of the connected quantum 4-pt function of Sec. \ref{subsec:4-pt_functions}. We start with the $\mathbf{k}$-space 4-pt function, given in Eq. (\ref{4-pt_func_k-space_Wightman}) as
\begin{equation}
    \label{4-pt_func_k-space_Wightman_again}
    \begin{split}
    \Tilde{G}^{(4)}_C(\eta,\{\mathbf{k}_i\})=&-6i\lambda\int^\eta_{-\infty} d\eta_z \frac{1}{(H\eta_z)^4}\\&
    \begin{split}
    \times\Bigg(&\Tilde{\Delta}^-(\eta_z,\eta,\mathbf{k}_1)\Tilde{\Delta}^-(\eta_z,\eta,\mathbf{k}_2)\Tilde{\Delta}^-(\eta_z,\eta,\mathbf{k}_3)\Tilde{\Delta}^-(\eta_z,\eta,\mathbf{k}_4)\\&-\Tilde{\Delta}^+(\eta_z,\eta,\mathbf{k}_1)\Tilde{\Delta}^+(\eta_z,\eta,\mathbf{k}_2) \Tilde{\Delta}^+(\eta_z,\eta,\mathbf{k}_3) \Tilde{\Delta}^+(\eta_z,\eta,\mathbf{k}_4)\Bigg).
    \end{split}
    \end{split}
\end{equation}
For the purposes of studying the general features of the IR limit, we will take $\mathbf{k}_i=\mathbf{k}$ $\forall i$. Using the $\mathbf{k}$-space Wightman function (\ref{k-space_2-pt_func}), the 4-pt function becomes\footnote{We will just use $\nu$ instead of $\nu_R$ here. As this is an $\mathcal{O}(\lambda)$ quantity, the non-trivial part of the renormalised mass won't feature.}
\begin{equation}
    \label{4-pt_func_Hankels}
    \begin{split}
        \Tilde{G}^{(4)}_C(\eta,\{\mathbf{k}_i\})=-6i\lambda\int^\eta_{-\infty} d\eta_z& \frac{1}{(H\eta_z)^4}\frac{\pi^4}{256H^4}(-H\eta)^{6}(-H\eta_z)^{6}\\&\times2i\Im\lsb\mathcal{H}_\nu^{(1)}(-k\eta)^4\mathcal{H}_\nu^{(2)}(-k\eta_z)^4\rsb.
    \end{split}
\end{equation}
Defining the quantities $K=-k\eta$ and $K_z=-k\eta_z$, the integral can be written as
\begin{equation}
    \label{4-pt_func_Hankels_K}
    \begin{split}
        \Tilde{G}^{(4)}_C(\eta,\{\mathbf{k}_i\})=\frac{3\lambda\pi^4 H^4(-\eta)^6}{64k^3}\int^\infty_K dK_z K_z^2\Im\lsb\mathcal{H}_\nu^{(1)}(K)^4\mathcal{H}_\nu^{(2)}(K_z)^4\rsb.
    \end{split}
\end{equation}
Since we are interested in the IR behaviour, we take the limit $K\ll1$ such that we can use the asymptotic behaviour of the Hankel functions
\begin{equation}
    \label{asymptotic_Hankel_func}
    \mathcal{H}_\nu^{(1)}(K)\simeq \frac{2^{-\nu}}{\Gamma(1+\nu)}K^{\nu}-i\frac{2^\nu\Gamma(\nu)}{\pi}K^{-\nu}
\end{equation}
such that
\begin{equation}
    \label{4-pt_func_Hankels_K_asymp}
    \begin{split}
        \Tilde{G}^{(4)}_C(\eta,\{\mathbf{k}_i\})=\frac{3\lambda\pi^4 H^4(-\eta)^6}{64k^3}\int^\infty_K dK_z K_z^2\Bigg[&\frac{2^{4\nu}\Gamma(\nu)^4}{\pi^4}K^{-4\nu}\Im\lsb\mathcal{H}_\nu^{(2)}(K_z)^4\rsb\\&+\frac{2^{2+2\nu}\Gamma(\nu)^3}{\pi^3\Gamma(1+\nu)}K^{-2\nu}\Re\lsb\mathcal{H}_\nu^{(2)}(K_z)^4\rsb\Bigg].
    \end{split}
\end{equation}
While this integral can't be computed in general, we can get some information about the IR behaviour of the 4-pt function. Consider the split of the integral
\begin{equation}
    \label{4-pt_func_Hankels_K_asymp_split}
    \begin{split}
        \Tilde{G}^{(4)}_C(\eta,\{\mathbf{k}_i\})=\frac{3\lambda\pi^4 H^4(-\eta)^6}{64k^3}\lb\int^\Lambda_K+\int_\Lambda^\infty\rb dK_z &K_z^2\Bigg[\frac{2^{4\nu}\Gamma(\nu)^4}{\pi^4}K^{-4\nu}\Im\lsb\mathcal{H}_\nu^{(2)}(K_z)^4\rsb\\&+\frac{2^{2+2\nu}\Gamma(\nu)^3}{\pi^3\Gamma(1+\nu)}K^{-2\nu}\Re\lsb\mathcal{H}_\nu^{(2)}(K_z)^4\rsb\Bigg],
    \end{split}
\end{equation}
for some parameter $\Lambda<1$. Focussing on the IR limit of the integral, $K<K_z<\Lambda$, we can take the limit $K_z<1$ such that we can use the approximate form of the Wightman functions (\ref{k_Wightman_IR_lim})
\begin{equation}
    \label{k_Wightman_IR_lim_again}
    \begin{split}
    \Tilde{\Delta}^\pm(\eta_z,\eta,\mathbf{k})\simeq \frac{\pi}{4Ha(\eta)^{3/2}a(\eta_z)^{3/2}}\Bigg[&\frac{4^{\nu}\Gamma(\nu)^2}{\pi^2}\lb\eta_z\eta\rb^{-\nu}k^{-2\nu}\\&\pm i\frac{1}{\pi\nu}\lb\lb\frac{\eta}{\eta_z}\rb^{\nu}-\lb\frac{\eta_z}{\eta}\rb^{\nu}\rb\Bigg].
    \end{split}
\end{equation}
Then, we can compute the IR region of the integral (\ref{4-pt_func_Hankels_K_asymp_split}) to find that the 4-pt function will have the following behaviour:
\begin{equation}
    \label{IR_behaviour_4-pt_k}
    \Tilde{G}^{(4)}_C(\eta,\{\mathbf{k}_i\})\sim K^{-6\nu}+K^{-3-4\nu}+K^{-3-2\nu}.
\end{equation}
The $K^{-6\nu}$ is just the term that we found in Eq. (\ref{k-space_4-pt_func}) and comes from the $K_z\rightarrow K$ limit. The other two contributions come from the limit $K_z\rightarrow\Lambda$. Converting these to coordinate space via a Fourier transform, one finds that
\begin{equation}
    \label{IR_behaviour_4-pt_x}
    G^{(4)}_C(\eta,\{\mathbf{x}_i\})\sim \abs{Ha(\eta)\mathbf{x}}^{-9+6\nu}+\abs{Ha(\eta)\mathbf{x}}^{-6+4\nu}+\abs{Ha(\eta)\mathbf{x}}^{-6+2\nu}.
\end{equation}
Since $\nu\leq 3/2$, it is immediately clear that the final term is subleading. However, for near-massless fields, $\nu\sim3/2$, the first and second terms give a similar contribution. As one increases the mass of the field, the second term is in fact the leading contribution over the first term. So, it appears that the contribution computed in Sec. \ref{subsec:4-pt_functions} is subleading. However, for this work, we only care about the 4-point function as a comparison tool between the stochastic theory and QFT so we can use this subleading term is it appears in the stochastic 4-point function. 

\section{Second-order stochastic parameters for matching with NLO term}
\label{app:stochastic_parameters_NLO}

For completeness, we will also include the matched stochastic parameters if we choose to reproduce the NLO term in the asymptotic expansion of the 2-point function in perturbative QFT. This choice doesn't make a difference to physical results. Repeating the procedure outlined in Sec. \ref{subsec:perturbative_stochastic_parameters}, one obtains the stochastic parameters to $\mathcal{O}(\lambda H^2/m^2)$ as
\begin{subequations}
    \label{O(lambdaH2m2_stochastic_parameters_NLO_match}
    \begin{align}
    m_S^2&=m^2_R+\frac{\lambda H^2}{8\pi^2}\lb-14+6\gamma_E-3\ln4+3\ln\lb\frac{M^2}{a(t)H^2}\rb\rb+\mathcal{O}\lb\lambda m_R^2\rb\\
    \sigma_{qq}^{2}&=\frac{H^3\alpha_R\nu_R}{4\pi^2\beta_R}\frac{\Gamma(2\nu_R)\Gamma(\frac{3}{2}-\nu_R)4^{\frac{3}{2}-\nu}}{\Gamma(\frac{1}{2}+\nu_R)}+\frac{\lambda H^5(-8+3\ln4)}{32\pi^4 m_R^2}+\mathcal{O}\lb\lambda H^3\rb,\\ \sigma_{qp}^{2}&=-\frac{51\lambda H^6}{32\pi^4m_R^2}+\mathcal{O}\lb\lambda H^4\rb,\\
    \sigma_{pp}^2&=\frac{H^5\beta_R^2\nu_R}{4\pi^2}\frac{\Gamma(-2\nu_R)\Gamma(\frac{3}{2}+\nu_R)4^{\frac{3}{2}+\nu_R}}{\Gamma(\frac{1}{2}-\nu_R)}-\frac{9\lambda H^7(-1+6\ln4)}{4\pi^4m^2_R}+\mathcal{O}(\lambda H^5).
    \end{align}
\end{subequations}


\bibliographystyle{hep}
\bibliography{main.bib}

\end{document}